\documentclass[twocolumn]{aa}

\usepackage{graphicx}
\usepackage{booktabs} 
\usepackage{natbib}
\usepackage{dsfont}
\usepackage{acronym}
\usepackage{physics}
\usepackage{dsfont}
\usepackage{siunitx}
\usepackage[dvipsnames]{xcolor}
\usepackage[colorlinks=true,linkcolor=blue, citecolor=blue, filecolor=blue, urlcolor=blue]{hyperref}
\usepackage{txfonts}

\acrodef{IFT}{information field theory}
\acrodef{PSF}{point spread function}
\acrodef{VI}{variational inference}
\acrodef{KL}{Kullback-Leibler divergence}
\acrodef{AGN}{active galactic nuclei}

\begin{document}

   \title{Latent-space Field Tension for Astrophysical Component Detection}
   \subtitle{An application to X-ray imaging}

   \author{Matteo Guardiani\inst{1,2}
          \and
          Vincent Eberle\inst{1,2}
          \and
          Margret Westerkamp\inst{1,2}
          \and
          Julian R\"ustig\inst{2, 5}
          \and
          Philipp Frank\inst{1,3}
          \and Torsten En{\ss}lin\inst{1,2,4,5}
          }

             
\institute{
  Max Planck Institute for Astrophysics, Karl-Schwarzschild-Straße 1, 85748 Garching, Germany\\
  \email{matteani@mpa-garching.mpg.de}
  \and
  Ludwig-Maximilians-Universität München, Geschwister-Scholl-Platz 1, 80539 Munich, Germany
  \and
  Kavli Institute for Particle Astrophysics \& Cosmology, Stanford University, Stanford, CA 94305, USA
  \and
  Excellence Cluster ORIGINS, Boltzmannstraße 2, 85748 Garching, Germany
  \and
  Deutsches Zentrum f\"ur Astrophysik, Postplatz 1, 02826, G\"orlitz, Germany
}

   \date{Received ; accepted }

 
  \abstract
  {
  Modern observatories are designed to deliver increasingly detailed views of astrophysical signals.
  To fully realize the potential of these observations, principled data-analysis methods are required to effectively separate and reconstruct the underlying astrophysical components from data corrupted by noise and instrumental effects.
  In this work, we introduce a novel multi-frequency Bayesian model of the sky emission field that leverages latent-space tension as an indicator of model misspecification, enabling automated separation of diffuse, point-like, and extended astrophysical emission components across wavelength bands.
  Deviations from latent-space prior expectations are used as diagnostics for model misspecification, thus systematically guiding the introduction of new sky components, such as point-like and extended sources.
  We demonstrate the effectiveness of this method on synthetic multi-frequency imaging data and apply it to observational X-ray data from the eROSITA Early Data Release (EDR) of the SN1987A region in the Large Magellanic Cloud (LMC).
  Our results highlight the method’s capability to reconstruct astrophysical components with high accuracy, achieving sub-pixel localization of point sources, robust separation of extended emission, and detailed uncertainty quantification.
  The developed methodology offers a general and well-founded framework applicable to a wide variety of astronomical datasets, and is therefore well suited to support the analysis needs of next-generation multi-wavelength and multi-messenger surveys.
  }

   \keywords{Methods: data analysis -- Techniques: image processing -- X-rays: general -- ISM: supernova remnants -- Galaxies: structure -- Techniques: photometric}

   \maketitle
%

\section{Introduction}

Astronomical observatories, however precise, inevitably imprint instrumental distortions onto the measured signals, corrupting the observational data.
These systematic distortions -- that stem both from the fundamental physics of measurement processes and from practical challenges inherent to observing faint and distant sources -- degrade the astrophysical signals of interest. 
While the construction of more powerful instruments can partly alleviate these issues, such endeavors demand substantial investments of time and resources \citep{ensslin_infogain}.
Moreover, the underlying challenge still remains: astronomical data typically contains more information than currently exploited by standard analysis methods.
Therefore, it is essential to develop advanced methodologies capable of maximally extracting the information encoded in available observations.

In particular, astronomical datasets commonly contain superpositions of multiple astrophysical components, including point sources (e.g. stars, quasars, supernovae), extended compact objects (nebulae, supernova remnants), diffuse emission (intracluster gas, Galactic cirrus), and observational backgrounds that are all blended together by instrument response functions. 
Accurately characterizing the physical properties of each component critically depends on our ability to separate and independently reconstruct them from observational data. 
This task, however, is complicated by shot noise, which is particularly severe in low-count X-ray and $\gamma$-ray regimes. Instrumental blurring effects, such as point spread functions (PSFs), add further complexity by mixing point sources and diffuse backgrounds.
Multi-frequency or multi-band observations exacerbate this complexity, as overlapping spectral features from distinct astrophysical phenomena become intertwined spatially and spectrally in each observed image.
Separating these mixed components to recover the underlying physical signals is an ill-posed inverse problem, yet it is crucial for interpreting multi-frequency survey data and identifying new astrophysical phenomena.

A number of methods have been developed to extract sources and separate components in astronomical imaging, each with important limitations. 
Heuristic source-finders like \texttt{SExtractor} \citep{sextractor, sextractor2} or \texttt{ProFound} \citep{profound} are widely used for their speed and simplicity. 
However they require manual parameter tuning, struggle to deal with complicated detector responses or noise, and provide limited uncertainty quantification.
While our application in this work focuses on X-ray observations, similar challenges in source separation and image reconstruction arise in other wavebands, such as radio or infrared.
In radio interferometry, the de facto standard deconvolution algorithm is \texttt{CLEAN} \citep{clean}, which iteratively subtracts PSF-scaled point sources. 
\texttt{CLEAN} effectively assumes all emission is point-like; as a result, extended and diffuse structures are poorly recovered. 
Multi-scale extensions of \texttt{CLEAN} \citep{ms_clean} (e.g. incorporating flux on multiple angular scales) offer improved recovery of diffuse features, but still often struggle with complex, multi-component scenes and require careful tuning of scale parameters. 

In recent years, machine learning methods have also been applied to component separation and source detection. 
Deep learning models (typically convolutional neural networks (CNNs)) can learn to identify and remove foreground sources from complex backgrounds given sufficient training data. 
For instance, \citet{deepnets} demonstrate a U-net based network that subtracts compact sources from Herschel far-infrared maps, significantly improving the analysis of extended emission in star-forming regions. 
Similarly, \citet{deepsource} use a CNN-based architecture to detect point sources in radio data and \citet{planck_ps} incorporate spectral training into their CNN architecture to detect point sources in microwave simulations.
While such approaches show promise in specific cases, they usually require supervised training with simulated data (e.g. injecting synthetic sources into real backgrounds as training pairs). 
This means the performance can be sensitive to the realism of the simulations and the method may not generalize outside the conditions it was trained on (different instruments, noise levels, or source populations). 
Moreover, purely data-driven networks do not inherently provide uncertainty estimates or physical interpretability, and they tend to treat instrument characteristics as fixed nuisances – often one must train a new model for each telescope or wavelength.

More sophisticated approaches recast component separation as a probabilistic inference problem, allowing a principled treatment of noise, instrument effects by incorporating prior knowledge. 
Early examples include maximum-likelihood or maximum-entropy reconstructions for high-energy instruments \citep{skilling1984, strong2003}, and sparse regularization techniques (wavelet thresholding, compressed sensing, Gaussian mixture models) to de-noise and deconvolve astronomical images \citep{wavelet_maxent, compressed_sensing, jolideco}.
These methods have seen success in specific contexts -- e.g. wavelet-based filtering has been used to detect faint sources in Planck and Fermi-LAT maps \citep{wavelet_ps, wavelet_planck} -- but they often require choosing a suitable basis or regularization parameter by hand.

Given the shortcomings of existing techniques, there is a need for a general and robust framework to automatically separate astrophysical components across multiple wavelengths, without extensive manual tuning or instrument-specific tailoring. Bayesian component separation methods aim to incorporate physical priors more transparently. For instance, in radio astronomy, \texttt{resolve} and \texttt{fast resolve} \citep{resolve, fast_resolve} apply hierarchical Bayesian modeling within the framework of \ac{IFT} \citep{ensslin_2009, ensslin2013information, ift2019} -- a Bayesian statistical field theory for signal inference -- to surpass the limitations of traditional deconvolution algorithms such as \texttt{CLEAN}. 
However, these methods currently struggle to model point-like sources.
In the high-energy regime, \citet{guglielmetti} applied a Bayesian forward modeling approach to separate diffuse and point-like components in ROSAT data, though their spline-based background model was limited to a small number of sampling points. A significant step forward was achieved by the \texttt{D3PO} algorithm (Denoising, Deconvolution, and Decomposition for Photon Observations; \citealt{d3po}), which leverages prior knowledge of the diffuse background's spatial correlations and the statistical distribution of point-source brightnesses. \texttt{D3PO} enables simultaneous denoising, PSF deconvolution, and component-wise reconstruction in X-ray and $\gamma$-ray observations.
Since then, similar techniques based on \ac{IFT} have been successfully employed to de-noise, deconvolve, and decompose the high-energy sky \citep{platzfermi, wmarg, eberle24arxiv}. 
In this work, we extend the \texttt{D3PO} framework and the broader IFT-based toolset by introducing a novel approach that achieves automated component detection and separation through latent-space field analysis.

Our method leverages the statistical information encoded in the latent random fields of hierarchical Bayesian forward models, while explicitly incorporating instrumental effects -- such as the point-spread function (PSF), exposure variations, and noise characteristics -- into a unified model of the observation. Within this framework, localized tension in the latent fields, defined as statistically significant deviations from their prior expectations, serves as a diagnostic indicator for the presence of astrophysical components.

Unlike traditional workflows that rely on sequential deconvolution, background subtraction, or source-detection stages, 
our approach jointly infers the most probable decomposition of the data into multiple astrophysical components, conditioned on the physical priors and the observational model. 
Because the instrument response is cleanly separated from the signal model, the method remains agnostic to the specific telescope or survey, and can be applied across different observational settings and wavelengths without retraining or tuning.

Crucially, the \ac{IFT} framework enables reconstruction of both the signal fields and their spatial and spectral correlation structures via hierarchical modeling of covariance hyperparameters. This allows the algorithm to learn, directly from the data, the morphology of diffuse emission (e.g., smoothly varying background versus small-scale fluctuations) as well as the spectral coherence of components across multiple bands. 
While the method introduced here is fully automated for point-like source detection, it also supports the identification and modeling of more extended or compact sources through the same latent-space diagnostics.

The proposed approach offers several key advantages.
Generality and scalability are achieved using fields, which are resolution independent and by formulating the inference through efficient solvers (building on the \texttt{NIFTy} library \citep{nifty, nifty5, nifty_re}) that can handle fairly large imaging data sets and arbitrarily gridded domains. 
The method natively supports multi-frequency (multi-band) data, solving for a consistent set of component maps across all input images -- this improves separation of components that are spectrally mixed in conventional analyses, and enables cross-band information transfer, e.g., reinforcing a source detection using its signal in another band with a higher signal-to-noise ratio (S/N).
By jointly modeling diffuse, point-like, and extended structures, the algorithm can disentangle overlapping sources of different morphology, yielding more accurate flux estimates and cleaner separation than single-technique methods. 
Another important benefit is in the low-count or low-surface-brightness regime: by combining data across frequencies and leveraging prior knowledge of correlation structures, the presented IFT-based reconstruction boosts the detectability of faint features that would be missed by threshold-based detectors. 
We demonstrate that our method produces de-noised, de-convolved, and decomposed images (see \citealt{eberle24arxiv}) that closely approximate the true sky components, while maintaining uncertainty quantification for each reconstructed field. In summary, this work introduces a powerful new tool for astrophysical imaging: a multi-component, multi-wavelength Bayesian reconstruction technique that separates superposed sources in a principled, instrument-agnostic way, pushing the state of the art in both sensitivity and fidelity of component recovery.

The rest of this work is structured as follows: in Section~\ref{sec:bayes}, we introduce the Bayesian inference framework and describe the hierarchical modeling techniques used for astrophysical component separation.
We begin with a one-dimensional toy model to illustrate the method’s core principles in Section~\ref{sec:1d-example} and then extend the approach to two-dimensional imaging data in Section~\ref{sec:imaging_example}.
Sections~\ref{sec:mf_model} and \ref{sec:mf_example} present the extension to multi-frequency data, including our prior modeling of diffuse, point-like, and extended sources across wavelength bands.
In Section~\ref{sec:erosita}, we apply our method to real SRG/eROSITA X-ray observations of the LMC SN1987A region, demonstrating its ability to reconstruct and disentangle complex astrophysical structures.
Finally, Section~\ref{sec:conclusions} summarizes our findings and outlines directions for future work.


\section{Methods \& Applications}
In order to accurately reconstruct astronomical signals from noisy and systematically biased data, 
we need to make assumptions on the properties of the signal and on the behavior of the observational instruments that collect the data.
In this work, we formulate these assumptions using the principles of Bayesian inference.
Specifically, since the signals we are interested in are well suited to be mathematically described by scalar or tensor fields,
we use the \ac{IFT} formalism.

\subsection{Bayesian astrophysical imaging}\label{sec:bayes}
The ultimate goal of astrophysical imaging is to infer the spatial and temporal distributions of physical observables of distant sources.
These observables include -- but are not limited to -- electromagnetic radiation from astrophysical sources, its polarization, and velocity fields.
\ac{IFT} allows to reconstruct the posterior distribution of high-dimensional and continuous fields from finite-dimensional data leveraging Bayes' theorem
\begin{equation}\label{eq:bayes}
	\mathcal{P}(s|d) = \frac{\mathcal{P}(d|s)\, \mathcal{P}(s)}{\mathcal{P}(d)},
\end{equation}
where we have indicated the signal field with $s$ and the data with $d$. 
We note that the signal field $s$ could depend on position, time, energy, polarization, and other physical dimensions, and may consist of multiple components.
\subsubsection{Sky brightness prior}
In the context of imaging, we can, for instance, choose to encode in the prior $\mathcal{P}(s)$ our assumptions about the surface brightness distribution across 
the observed patch of sky. In this work, we address the problem of designing a prior for the sky brightness distribution across different wavelengths.
Following the approach outlined by \cite{wmarg,eberle24arxiv}, we formulate these priors in the language of Bayesian hierarchical forward models. 

To provide a standardized interface for model implementation and inference, we formulate our models as generative models. 
These define a mapping from a latent Hilbert space $ \Omega_\xi$  to a target space $\Omega_s$, where the signal $s$ is defined.
For simplicity, we assume that the prior on the latent-space variables in $\Omega_\xi$ follows a multivariate standard normal distribution.
The generative mapping is then constructed to ensure that the transformed variables in $\Omega_s$ adhere to the desired distribution $\mathcal{P}(s)$.
We note that in practice $\Omega_s$ is a finite-dimensional parametrization of the more general Hilbert space in which the theoretical signal field $s$ resides.  
This enables practical inference while maintaining a well-defined connection to the underlying infinite-dimensional problem.

Specifically, this work addresses the problem of detecting, separating, and reconstructing astrophysical signal components.  
Because the observed data consists of a noisy and systematically biased superposition of multiple components (e.g., from different astrophysical sources), reconstructing and separating them is an inherently ill-posed problem.  
That is, an infinite number of signal field configurations can explain the same data realization.  
As a result, the component separation task becomes entirely prior-driven, since the data alone cannot discriminate between different plausible solutions.
This motivates the need for prior models that better encode our knowledge of complex astrophysical signal distributions.

\subsubsection{Likelihood}
Having introduced the role of the prior $\mathcal{P}(s)$ for a given signal of interest  $s$, we now turn to the likelihood $\mathcal{P}(d|s)$, which connects the signal to the observed data $d$.
In this work, we define the signal space to encompass physically plausible configurations of the observed astrophysical fields, making the prior independent of any instrumental effects introduced during measurement.
To model how the signal $s$ gives rise to the observed data  $d$, we account for the instrument’s response function,
\begin{equation*}
	R(s) \coloneqq \expval{d}_{\qty(d|s)},
\end{equation*}
which describes the expected data measurement corresponding to a given signal configuration under the assumption that the noise has zero mean.
This involves constructing a forward model that accurately represents how the signal is processed by the instrument.
For efficient signal reconstruction, the response model must be both computationally scalable and mathematically differentiable.
We denote the instrument's response operator with $R$. 
For X-ray observatories, $R$ is typically approximated by chaining three operators: the \ac{PSF} operator $O$ to an exposure operator $E$ and a masking operator $M$, as discussed in \cite{eberle24arxiv},
\begin{equation*}
	R = M \circ E \circ O.
\end{equation*}
While the \ac{PSF} represents the conditional probability distribution of measuring a photon at a specific location across the field of view given its original location, the exposure encodes the observation time, the effective area and vignetting, and the masking 
operator $M$ accounts for detector characteristics, such as dead pixels or regions obscured by hardware components.
In practice, additional effects such as pile-up, charge-coupled device (CCD) contamination, and energy-dependent response functions may influence the instrument’s behavior. 
Including these effects in the response model $R$ can improve the accuracy of the signal reconstruction $s$. 
However, since this work focuses on developing efficient prior models, and given the trade-off between model complexity and computational cost, we restrict our analysis to the aforementioned systematics.
It is important to note that the response model $R$ is independent of the signal $s$.
This independence ensures that any prior model developed for $s$ can, in principle, be applied to data from any instrument.
To implement the astrophysical response $R$ into our forward models, we make use of the universal Bayesian imaging kit (\texttt{J-UBIK}; \cite{jubik}).
For photon-count instruments like eROSITA, the likelihood of a specific number of counts $d_{i}$ in a pixel $i\in \qty{1, \dots, N}$ given an observed rate $\lambda_{i} = \qty(R(s))_{i}$ is given by the Poisson distribution
\begin{equation*}
	\mathcal{P}(d_{i}|s) = \frac{\lambda_{i}^{{d_{i}}}}{d_{i}!}\, e^{-\lambda_{i}}.
\end{equation*}
Finally, since -- conditional to a fixed signal -- the number of recorded photons in each pixel $i$ is assumed to be statistically independent from each other, we can write the combined likelihood of all pixels as
\begin{equation}
	\mathcal{P}(d|s) = \prod_{i=1}^{N} \mathcal{P}(d_{i}|s) = \prod_{i=1}^{N} \frac{\lambda_{i}^{{d_{i}}}}{d_{i}!}\, e^{-\lambda_{i}}.
\end{equation}
For practical purposes, we work with the so-called information Hamiltonian of the likelihood, defined as
\begin{equation*}
\mathcal{H}(d|s) \coloneqq - \log{\mathcal{P}(d|s)} = \sum_{i=1}^{N} \qty[ \lambda_{i}(s) - d_{i} \log{\lambda_{i}(s)} + \log{d_{i}!} ],
\end{equation*}
where the final term, $\log{d_{i}!}$, is independent of the signal $s$ and can therefore be neglected during inference.
This formulation is proportional to the well-known Cash statistic \citep{cash}, commonly used to deal with Poisson-distributed data from astrophysical observations.

\subsubsection{Posterior}
Once the prior and likelihood distributions for the fields of interest are specified, Bayes’ theorem for fields (Eq.~\eqref{eq:bayes}) can be used to determine the posterior probability distribution of the signal $s$, 
conditioned on the observation $d$. However, directly evaluating the posterior can be computationally prohibitive.
The main challenge lies in calculating the evidence term
\begin{equation*}
\mathcal{P}(d) = \int_{\Omega_s} \mathcal{P}(d|\tilde{s})\, \mathcal{P}(\tilde{s})\, \mathcal{D}\tilde{s},
\end{equation*}
which requires integrating over the parameter space $\Omega_s$ of the signal field $s$. For the astrophysical applications discussed in this work, $\Omega_s$ is multi-million dimensional, making the evaluation of this integral infeasible with direct methods.
To overcome this computational barrier, we employ \ac{VI}, which approximates the posterior distribution efficiently and scales to high-dimensional parameter spaces.
In \ac{VI}, rather than directly evaluating the evidence $\mathcal{P}(d)$, we approximate the posterior distribution using a family of distributions $\mathcal{Q}_\alpha(s|d)$, parameterized by the variational parameters $\alpha$. 
This approximation is achieved by minimizing the \ac{KL}
\begin{equation*}
	\mathcal{D}_\text{KL}\qty(\mathcal{Q_\alpha}||\mathcal{P}) \coloneqq \int_{\Omega_s} \mathcal{Q}_\alpha\qty(\tilde{s}|d)\, \log{\frac{\mathcal{Q}_\alpha\qty(\tilde{s}|d)}{\mathcal{P}\qty(\tilde{s}|d)}}\, \mathcal{D}\tilde{s},
\end{equation*}
with respect to $\alpha$. 
In this work, we approximate the posterior distribution $\mathcal{Q}_\alpha(s|d)$ using geometric variational inference (geoVI, \cite{Frank_2021}). 
Following metric Gaussian variational inference (MGVI, \cite{mgvi}), geoVI extends traditional variational inference by leveraging the Fisher information metric to better capture the local geometry of the posterior distribution.
The Fisher metric reflects the curvature of both the likelihood and prior, providing a natural measure of the posterior’s geometric structure. By utilizing this metric, geoVI constructs a local isometry -- a transformation that maps the curved parameter space 
to a Euclidean space while preserving its intrinsic geometric properties. In this transformed space, the posterior distribution is closer to Gaussian, enabling a more accurate Gaussian variational approximation.
This approach allows geoVI to represent non-Gaussian posteriors with higher fidelity compared to MGVI, improving the quality of inference for complex Bayesian models. 
The ability to exploit the posterior’s geometry is particularly valuable for the astrophysical component separation problem addressed in this work, for which accurate modeling and inference are critical.

\subsection{A one-dimensional example}\label{sec:1d-example}
\begin{figure*}[!htbp]
   \centering
   \includegraphics[width=1.\textwidth]{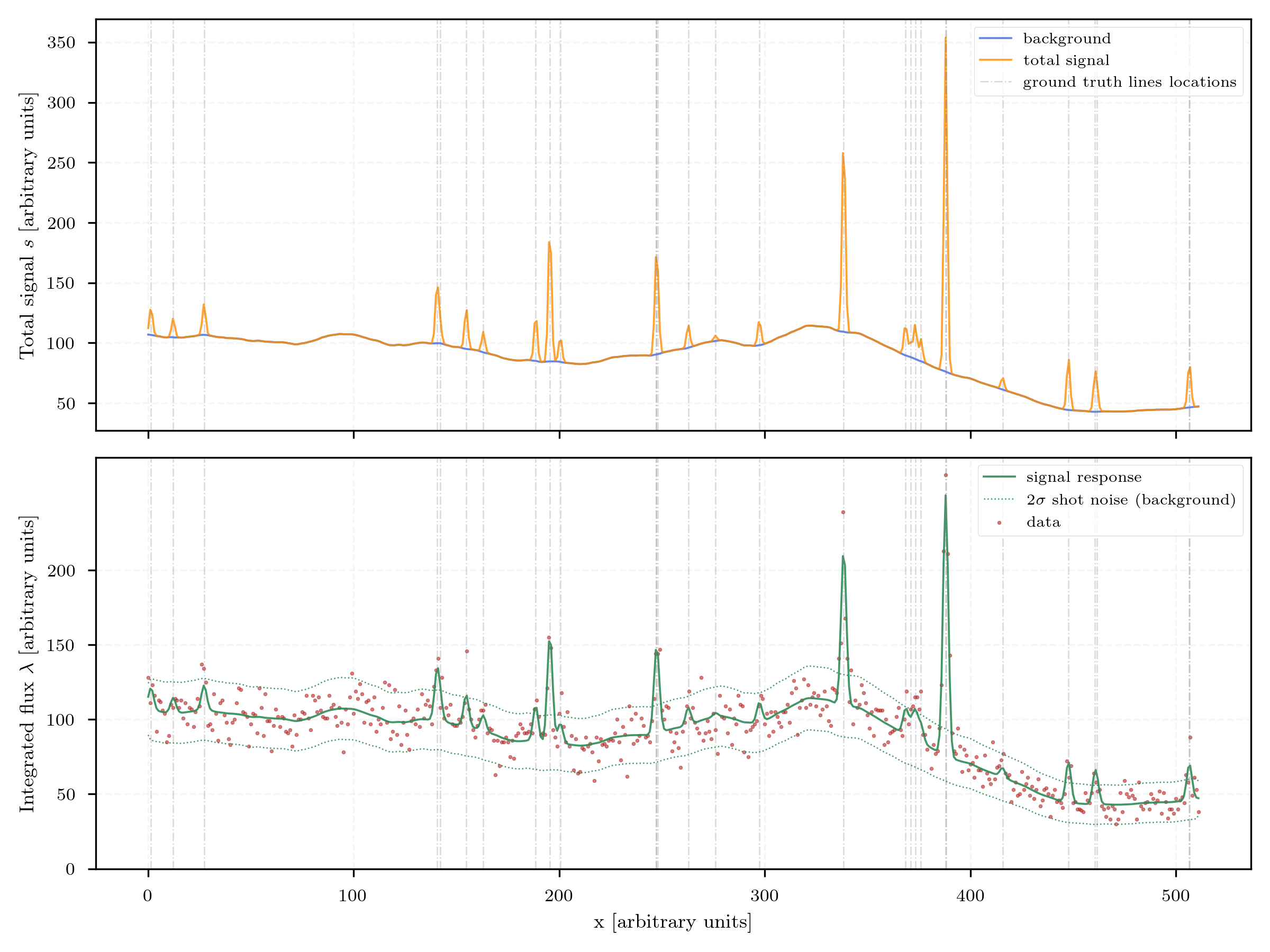}
      \caption{1D example of the automatic component modeling method. 
      \textit{The top panel} displays the background $b$ (in blue) and the total signal $s$ (in orange) which is the superposition of the background $b$ and the foreground line $f$ signal components.
      \textit{The bottom panel} shows the signal response (solid green line), obtained by convolving the total signal $s$ with a Gaussian \ac{PSF}. 
      Additionally, it displays a synthetic data realization of the signal response (in red), along with the $2 \sigma$ shot noise level contours associated with the background (dotted green lines).
      In both panels the vertical dashed-dotted gray lines represent the ground truth line component $f$ locations.
      We note that a large fraction of the line signal $f$ is buried beneath the background noise level.
      }
         \label{fig:1d-example}
\end{figure*}

In this work, we aim to exploit latent-space information to improve component separation and detect faint foreground structures from noisy data.
To illustrate the core ideas, we begin with a simple one-dimensional example.
Assume the total signal field of interest $s(x)$ consists of a positive, diffuse, correlated background component $b(x)$, and a localized, uncorrelated line component $f(x)$
\begin{equation*}
	s(x) = b(x) + f(x).
\end{equation*}
The task is to reconstruct the total signal $s$ and, importantly, separate the background emission $b$ from the line component $f$ from noisy data.
We show an example of this task in Fig.~\ref{fig:1d-example}.

To model the background signal $b(x)$ we use a lognormal process
\begin{equation}\label{eq:lognormal_field}
	b(x) = e^{\tau(x)},
\end{equation}
where $\tau(x)$ is the Gaussian process
\begin{equation*}
	\tau \sim p(\tau|T) = \mathcal{G}(\tau, T) \coloneqq \frac{1}{\sqrt{\qty|2\pi T|}}\, \exp\qty(-\frac{1}{2} \tau^\dagger T^{-1}\tau),
\end{equation*}
with covariance $T$.
The lognormal distribution of $b$ ensures positivity and naturally captures variations on logarithmic scales, which are characteristic of astrophysical emission signals.
Following \cite{arras_cf}, and assuming prior spatial homogeneity and isotropy, we adopt a flexible yet computationally efficient parametrization of the covariance $T$ in harmonic space, known as the ``correlated field model''.
The generative model for $\tau(x)$ can be written as
\begin{equation}\label{eq:1d-xi}
	\tau_x = \mathcal{F}_{xk} \sqrt{\hat{T}_{kk'}}\,  \xi_{k'},
\end{equation}
where $\xi_k$ denotes the latent-space excitations of the Gaussian process in harmonic space $\Omega_k$, indexed by modes $k \in \Omega_k$, and $\hat{T}$ is the Fourier transform of the covariance $T$.
In Eq.~(\ref{eq:1d-xi}), we have employed Einstein’s summation/integration convention over repeated indices, which we will continue to adopt throughout this work.
The Fourier transform operator $\mathcal{F}_{xk}$ establishes the connection between field's support in position space $\Omega_x$ and in harmonic space $\Omega_k$.
More generally, for a field defined on vector spaces with coordinates $\vb{x}$ and $\vb{k}$, the (inverse) Fourier transform operator acts as
\begin{equation*}
	\mathcal{F}_{\vb{x}\vb{k}} \tau_{\vb{k}} \coloneqq \frac{1}{(2\pi)^D} \int_{\Omega_k} e^{i\vb{k} \cdot \vb{x}}\, \tau_{\vb{k}}  \dd{\vb{k}} = \tau_{\vb{x}},
\end{equation*}
where $D$ is the number of dimensions of $\Omega_x$ and $\Omega_k$.
It maps fields defined in the spatial domain to their representation in the frequency domain, enabling efficient manipulation and analysis of spatial correlations, such as those encoded in the covariance structure $T$.
In particular, the assumption of statistical homogeneity and isotropy diagonalizes the Fourier-transformed covariance
\begin{equation*}
      \hat{T}_{kk'} = \qty(2\pi)^D\, \delta\qty(\vb{k}-\vb{k}')\, P_T(\abs{\vb{k}}),
\end{equation*}
where $P_T(\abs{\vb{k}})$ is the prior isotropic power spectrum of the process.
In our standardized generative models, the latent-space excitations $\vb*{\xi}$ are standard-normally distributed
\begin{equation*}
	\vb*{\xi} \sim \mathcal{P}(\vb*{\xi}) = \mathcal{G}(\vb*{\xi}, \mathds{1}).
\end{equation*}

To model the line signal $f(x)$ we use a sum of Gaussian profiles centered at $N$ uniformly sampled $x_i \in \Omega_x$ locations (in the example in Fig.~\ref{fig:1d-example}, $N=30$)
\begin{equation*}
	f(x) = \sum_{i=1}^N \frac{f_i}{\sqrt{2\pi}\, \sigma_f}\,  \exp\qty(-\frac{(x-x_i)^2}{2 \sigma_f^2})
\end{equation*}
with fixed variance $\sigma_f^2$ and intensity $f_i$ sampled from an inverse-Gamma distribution
\begin{equation*}
	f_i \sim \mathcal{P}(f_i | \alpha, q) = \frac{q^\alpha}{\Gamma(\alpha)}\, f_{i}^{-\alpha-1} e^{-\frac{q}{f_{i}}},
\end{equation*}
where $\alpha$ is known as the shape parameter and $q$ controls the scale of the intensity.

To mimic a very rudimentary astronomical response, the total signal $s$ is then convolved with a Gaussian \ac{PSF} and multiplied by an exposure time $E$, constant across the spatial dimension.
We refer to the resulting integrated flux, from which the Poisson counts are drawn, as the signal response, shown in Fig.~\ref{fig:1d-example} together with the synthetic data realization.

Separating the two components is challenging, as the data only constrains their sum. 
This task is further complicated by noise and the unknown nature of the line component $f$. 
In particular, neither the number nor the locations of the lines are known a priori, and many may lie below the noise level, making them effectively indistinguishable from the diffuse background $b$.
In order to circumvent this problem, we need to introduce missing information encoded as prior knowledge.
Specifically, we impose smoothness assumptions on the background component through a prior on the power spectrum of the Gaussian field $\tau$.
This assumption is crucial for separating $f$ from $b$, since the presence of sharp, localized lines explicitly violates the prior expectation of smoothness.

\begin{figure*}[!htbp]
   \centering
   \includegraphics[width=1.\textwidth]{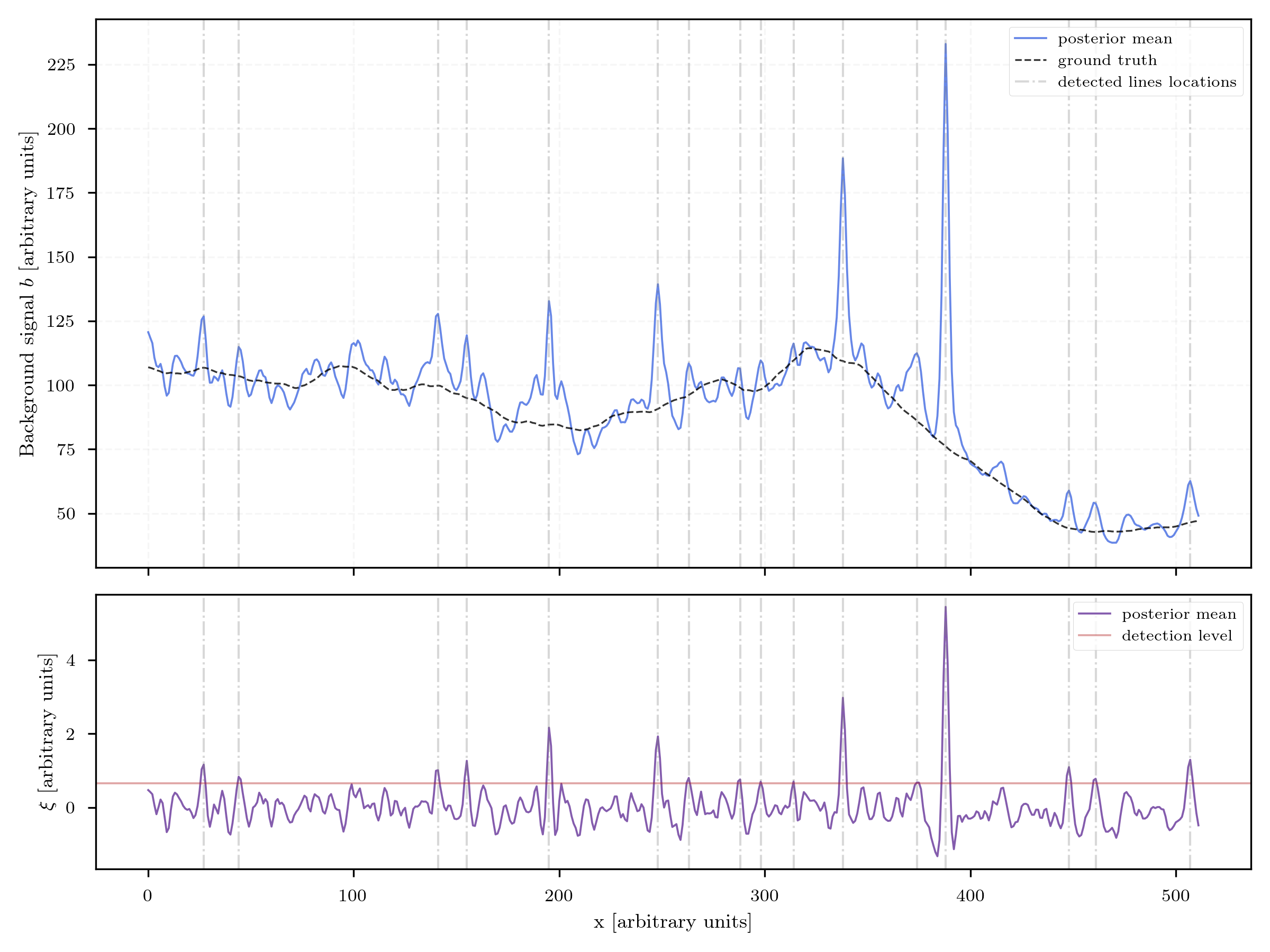}
   \caption{Background-only reconstruction.
      \textit{The top panel} displays the posterior mean of the background-only reconstruction (solid blue line) in comparison to the ground truth background signal response (dashed black line).
      \textit{The bottom panel} shows corresponding latent position-space excitations $\xi$ (purple solid line). The detection level for the line component was set to $\sigma_\text{thresh} = 0.6$ in prior units and it is displayed in red (solid horizontal line).
      The detected line positions for the given threshold $\sigma_\text{thresh}$ are displayed as dashed vertical gray lines in both panels.}
         \label{fig:1d-xis}
\end{figure*}

To begin reconstruction, we first infer the signal assuming it is fully described by the diffuse background model $b = e^\tau$ alone.
Learning $b$ implicitly involves inferring its correlation structure, i.e., the power spectrum of $\tau$ \citep{arras_cf}.
However, if the data contains signatures of unresolved or point-like features, the background model -- built on the assumption of homogeneity and isotropy -- will absorb these features into the continuous background field.
This will have an impact on the inferred correlation structure.
As a result, the posterior correlation structure will no longer reflect the original prior assumptions, but instead encode inhomogeneities and anisotropies introduced by the unmodeled foreground components.
This leads to a degraded fit: the reconstructed background appears artificially rough, as it tries to explain both the diffuse and line-like components under the assumption of smoothness.
The result is a contamination of the inferred power spectrum by white-noise-like features, typically visible as a flattening or kink at higher $k$ modes.
We illustrate such a background-only fit in Fig.~\ref{fig:1d-xis}. 

While this fit may appear unsatisfactory due to the issues outlined above, it provides a powerful insight: by analyzing the latent excitation field $\xi$, we can detect where the current model assumptions (a single background with long-ranged correlations) break down.
Recall that the log-normal field $b(x) = e^{\tau(x)}$ is constructed via a Gaussian process, which in turn is the convolution of white Gaussian excitations $\xi_x$ with a correlation kernel defined by $\sqrt{T}$.
If the data conforms well to the prior -- i.e., if the signal correlation structure was perfectly represented by the kernel -- these latent-space excitations should remain close to zero, with a variance of one.
However, if the data includes features that are incompatible with the assumed smoothness -- such as uncorrelated line emission -- the excitations will grow in magnitude at those locations.
Thus, the excitation field $\xi$ contains implicit information about violations of the background prior assumptions.
In this way, we can use the results of a background-only model fit as a diagnostic in the latent space to identify and localize the line component $f$.
Specifically, since we are interested in finding the locations at which possible model misspecification might happen, we look at the excitations in position space, i.e., $\xi_x = \mathcal{F}_{xk}\, \xi_k$.
Keeping in mind that the excitations are a-priori standard normally distributed, we can define a detection threshold $\sigma_\text{thresh}$ for the excitations in position space above which we consider a possible line detection.
We present a deeper theoretical analysis of the latent-space representation of the signal in Appendix~\ref{app:1d_ps_theory}.
\begin{figure*}[!htbp]
   \centering
   \includegraphics[width=1.\textwidth]{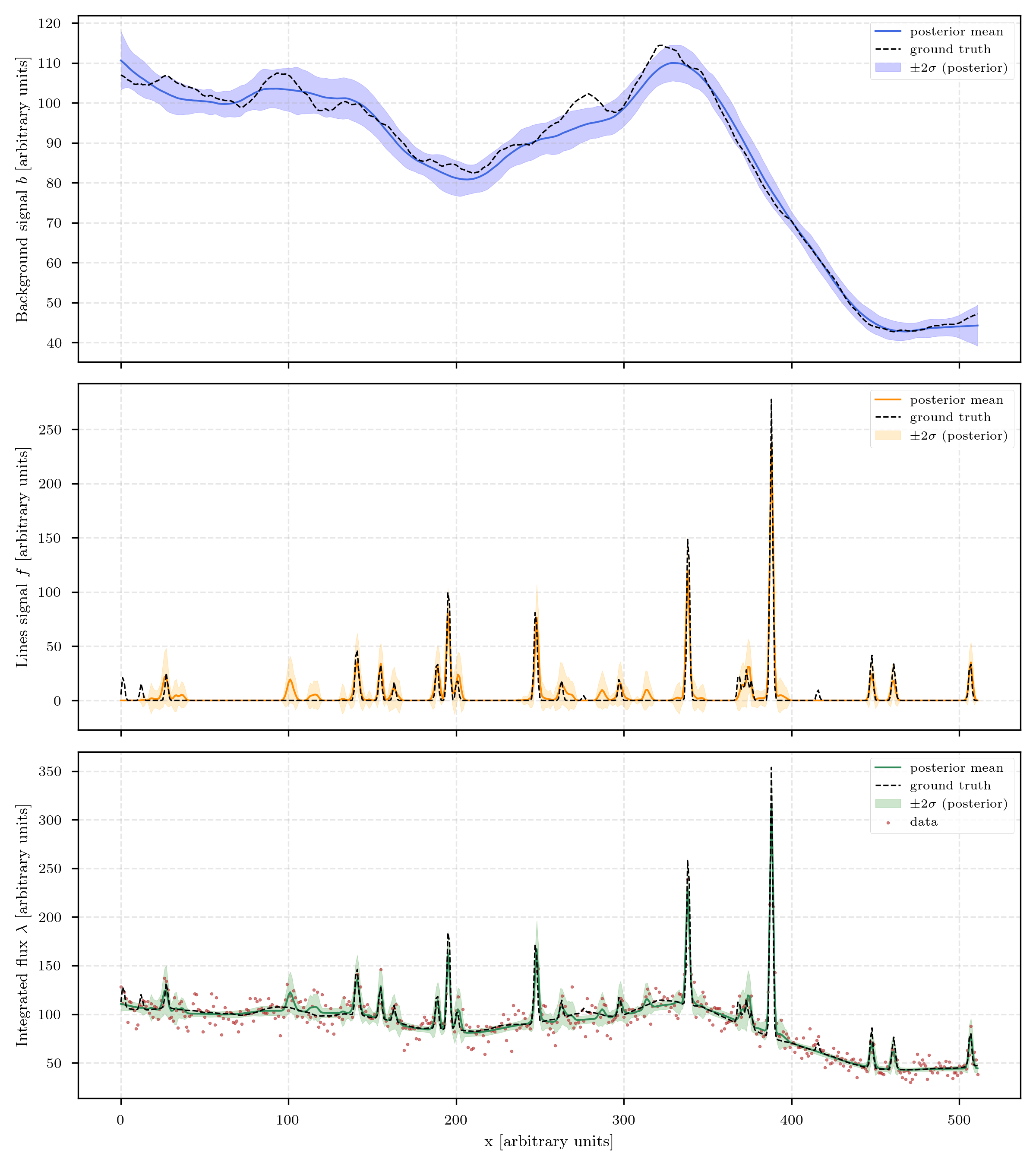}
   \caption{Results of one-dimensional automatic component modeling method.
      \textit{The top panel} displays the posterior mean of the background component $b$ (solid blue line) in comparison to the ground truth background signal (dashed black line).
      \textit{The central panel} depicts the posterior mean of the line component $f$ (solid orange line) in comparison to the ground truth line signal (dashed black line).
      \textit{The bottom panel} shows the posterior mean of the total signal $s$ (solid green line) in comparison to the ground truth signal (dashed black line). The data is also shown in red.
	Shaded regions denote the $\pm 2\sigma$ posterior contours for each component in the respective panels.}
         \label{fig:1d-results}
\end{figure*}
After identifying potential model misspecification locations, we iteratively refine the model by introducing line components centered at these positions.
The validity of the extended model can be assessed by monitoring the evidence lower bound (ELBO), which is computed using \texttt{NIFTy} and provides a principled criterion to determine whether the inclusion of additional components improves the overall fit.
This procedure can be repeated until no further excitations exceed the detection threshold.
The result of this process is shown in Fig.~\ref{fig:1d-results}.

While lowering the detection threshold may help recover additional true line components, it also increases the likelihood of introducing spurious ones in regions without actual signal.
Such false positives typically exhibit lower inferred intensity and larger posterior uncertainty, allowing them to be filtered out in a post-processing step.
The efficiency of this detection process can be significantly enhanced by analyzing the response of the correlated field to a line-like perturbation in the data.
This is discussed in Appendix~\ref{app:1d_ps_theory}, with a practical demonstration provided in Sec.~\ref{sec:erosita}.

Overall, we find that the final reconstruction not only closely matches the ground truth signal across nearly all locations, but also provides a clear separation between the background component $b$ and the line component $f$.
Moreover, with the chosen detection threshold, our method successfully recovers point sources below the $2\sigma$ background shot-noise level.

\subsection{An imaging example}\label{sec:imaging_example}
We now want to apply the same technique to imaging. 
Specifically, we focus on the task of separating diffuse and point-like emission, crucial for astrophysical imaging.
We start by showing how to use the method on a synthetic-data example.
\subsubsection{Imaging setup}\label{sec:imaging_setup}
\begin{figure*}[!htbp]
   \centering
   \includegraphics[width=1.\textwidth]{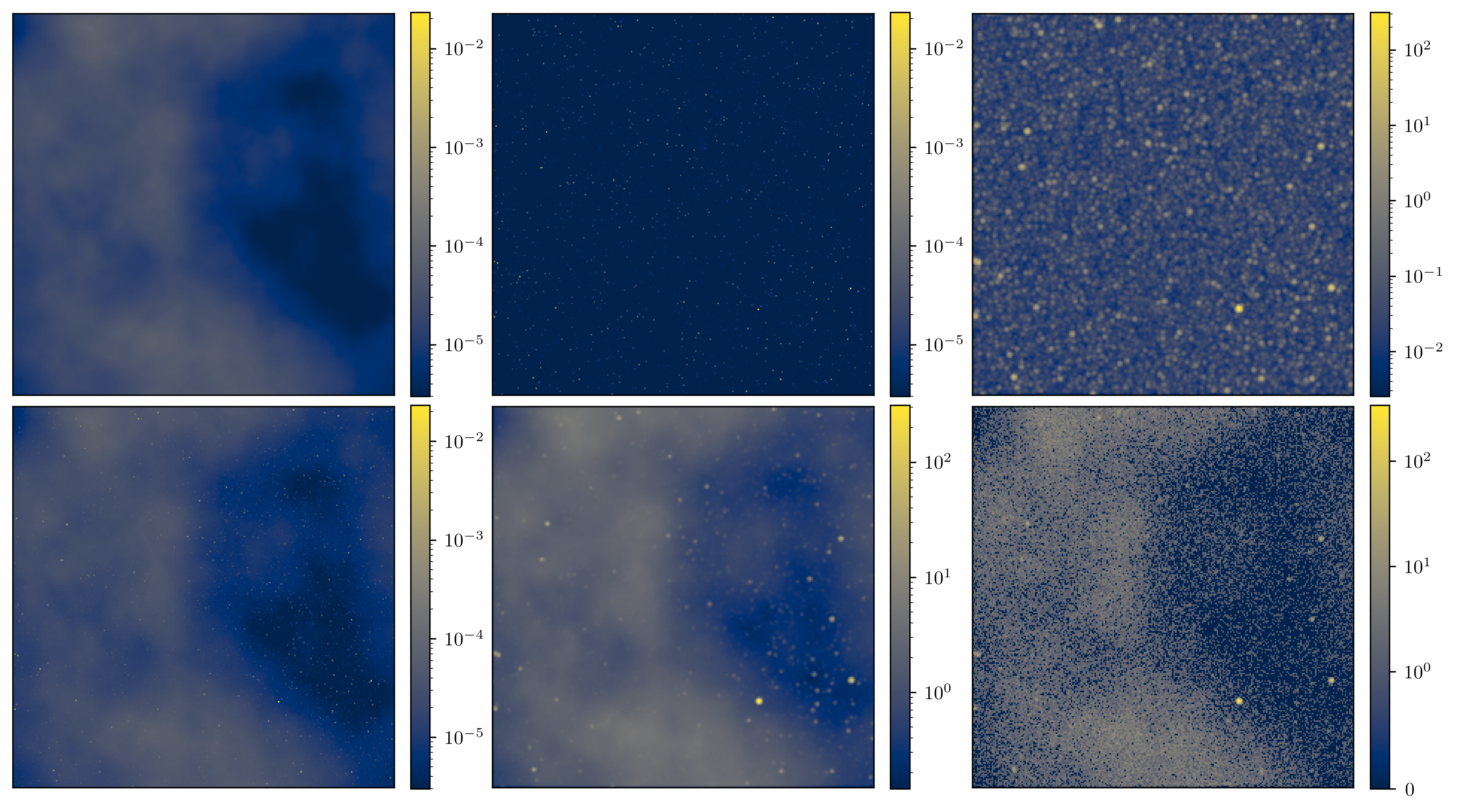}
   \caption{
   Setup of the synthetic component separation problem for imaging.
   \textit{The top row} displays a realization of the diffuse emission component (left), the point-source component (middle), and the convolved point-source component (right), which has undergone Gaussian \ac{PSF} convolution and multiplication by a spatially constant exposure of $\SI{80000}{\second}$.
   \textit{The bottom row} presents the corresponding sky emission, obtained as the sum of the diffuse and point-source components (left), the convolved and exposure-multiplied sky emission (middle), and data generated by a Poisson realization of the observed simulated sky (right).
   }
         \label{fig:2d-setup}
\end{figure*}

For instance, we consider the synthetic example shown in Fig.~\ref{fig:2d-setup}.
This example should entail all the main ingredients of an astrophysical imaging task with count data.
For the diffuse emission field model $b(\vb{x})$, we use a two-dimensional extension of the lognormal correlated field model from Eq.~(\ref{eq:lognormal_field}), following the approach used in real imaging scenarios by \citet{arras_cf, platzfermi}.
For the point-source emission field model we use a product of inverse-Gamma distributions
\begin{equation*}
	\mathcal{P}(f | \alpha, q) = \prod_{x_i}^N \frac{q^\alpha}{\Gamma(\alpha)}\, f_{x_i}^{-\alpha-1} e^{-\frac{q}{f_{x_i}}},
\end{equation*}
where $x_i \in \qty{0, \dots, N}$ represent the regular grid points onto which the point-source field $f$ is discretized, with $N$ being the total number of pixels.
We note that the inverse-Gamma prior can be made resolution-independent for $\alpha = 1.5$, since coarsening the resolution of the point-source field in this case still yields a field that is inverse-Gamma distributed, albeit with rescaled parameters\footnote{This property was discussed by \citet{d3po} and \citet{guglielmetti}, and rigorously analyzed by \citet{Girón2001}, where it is shown that under certain scaling relations, the inverse-Gamma distribution is stable under discretization changes.}.
Following the one-dimensional example from Sec.~\ref{sec:1d-example}, we construct the total photon flux by summing the PSF-convolved point-source and diffuse fields, and multiplying the result by a constant exposure time of $E = \SI{80}{\kilo\second}$, assumed for simplicity to be uniform across the field of view.
From this resulting flux field, we generate a Poissonian realization, which serves as the synthetic data shown in Fig.~\ref{fig:2d-setup}.

\subsubsection{Imaging example inference}
Our goal is to perform inference in this two-dimensional setting in a manner analogous to Sec.\ref{sec:1d-example}.
As discussed in Sec.\ref{sec:imaging_setup}, we assume in this example that each pixel contains both a point-source component and a diffuse emission component.
This assumption significantly increases the complexity of the inference problem, as the optimization must now determine how to attribute the flux in each pixel: whether to the point-source or the background component.

In real astronomical data, we do not know a priori whether a point source is present in every pixel. 
Moreover, the true positions of point sources rarely align with pixel centers.
To account for this, we model the point-source field in continuous space, representing each source explicitly by its position and flux.
The point-source field is then given by
\begin{equation}\label{eq:point-source-model}
	f(\vb{x}) = \sum_{i=0}^{N_\text{ps}} \delta(\vb{x}-\vb{x}_i)\, f_i,
\end{equation}
where $\vb{x}$ denotes the position of the $i$-th source and $f_i$ its corresponding flux, with $i \in \{0, \dots, N_\text{ps}\}$.
Because $\vb{x}$ lies in continuous space, this representation is not restricted to pixel centers, enabling sub-pixel localization of sources from the data.
A straightforward way to include this model in the likelihood would be to evaluate the \ac{PSF} at each position $\vb{x}$ and scale it by $f_i$.
However, this approach can become computationally expensive -- especially when the \ac{PSF} varies significantly across the field of view \citep{jubik, veberle_psf} -- as it requires evaluating the \ac{PSF} at updated positions during inference.

\begin{figure*}[!htbp]
   \centering
   \includegraphics[width=1.\textwidth]{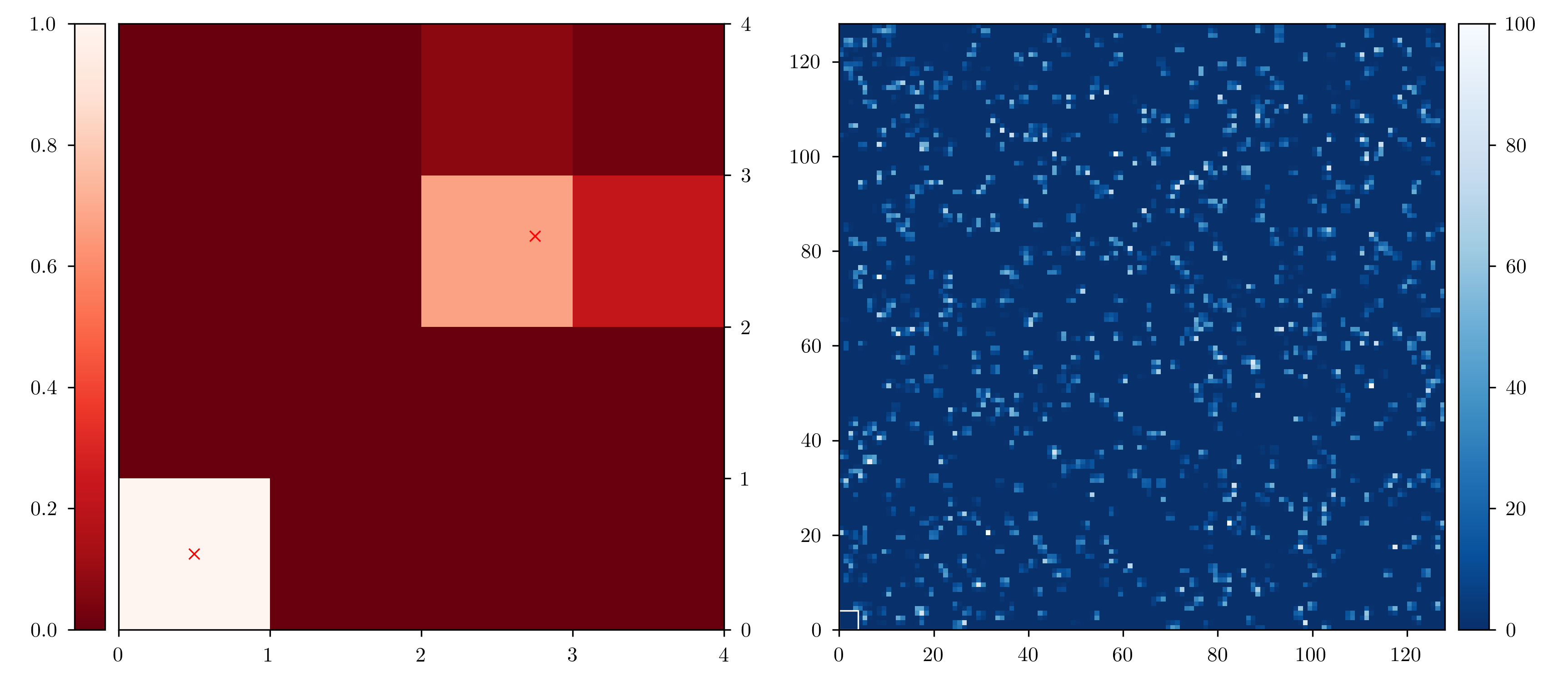}
   \caption{Interpolation of the point-source field flux onto a regular pixel grid.
   \textit{The left panel} shows the interpolation result for two, unit-flux point sources located at $\qty(0.5, 0.5)$ (center of the pixel) and at $\qty(2.75, 2.6)$ on a $4\times4$ pixel grid. 
   The location of the point sources is marked with a red cross. 
   \textit{The right panel} displays a random realization of $1000$ uniformly distributed point sources on a $128\times128$ pixel grid with flux values uniformly distributed between $0$ and $100$. A white rectangle in the lower-left corner indicates a region of the same size as that shown in the left panel, providing a spatial scale reference.
   }
         \label{fig:ps-interpolation}
\end{figure*}

To mitigate this challenge, we adopt an alternative but equivalent strategy.
Instead of evaluating the \ac{PSF} directly at every source position, we bilinearly interpolate the point-source fluxes onto a regular pixel grid and subsequently convolve the resulting image with the local \ac{PSF}.
This procedure is equivalent in the case of spatially invariant \acp{PSF}, provided the \ac{PSF} does not vary significantly within a pixel.
For spatially varying \acp{PSF}, the situation is more complex, as the PSF itself may already be interpolated from a finite sampling \citep{veberle_psf}.
Nonetheless, our approach retains sub-pixel accuracy in source positioning and enables efficient convolution, substantially accelerating the likelihood evaluation.
Figure~\ref{fig:ps-interpolation} shows both an example of the interpolation procedure and a random realization of $100$ point sources with uniformly distributed brightness, illustrating the structure of the prior over point-source fields under this model.

\begin{figure*}[!htbp]
   \centering
   \includegraphics[width=1.\textwidth]{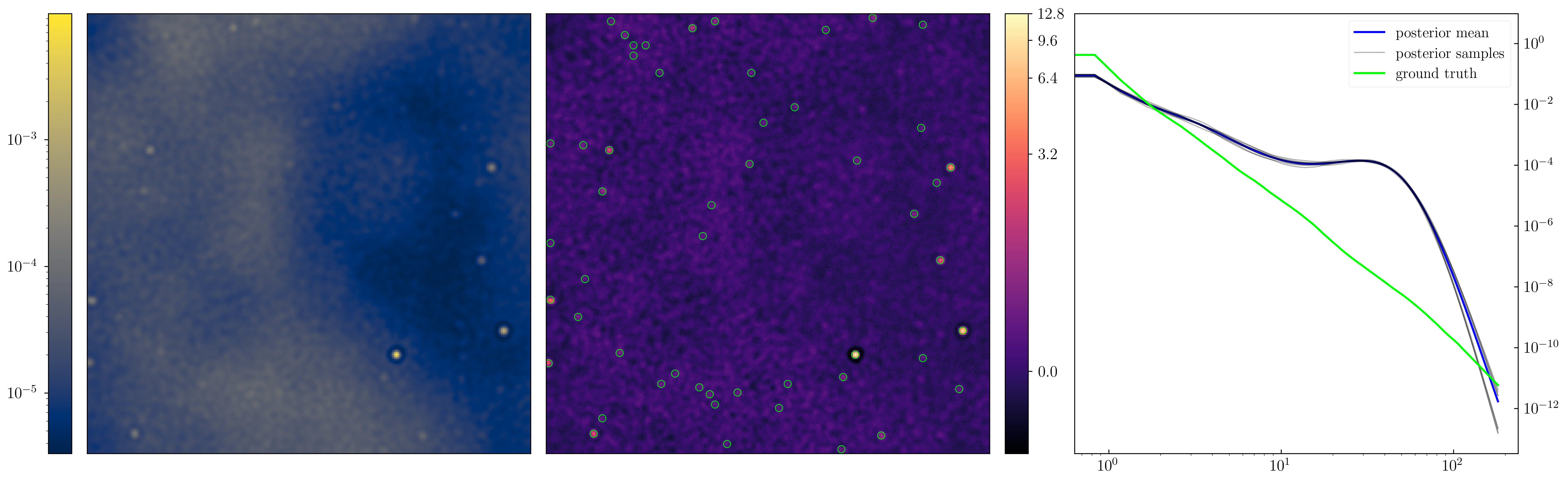}
   \caption{Result of background-only reconstruction.
   \textit{Left:} posterior mean of the reconstructed background field after two geoVI iterations. Artifacts at the point-source length scale are visible (low-flux coronas around the point sources, wiggles in the diffuse field).
   \textit{Center:} position-space excitation field $\xi$ after two geoVI iterations. Detected point-source candidates above the $\xi = 0.65$ threshold are circled in lime.
   \textit{Right:} reconstructed spatial power spectrum. The posterior mean is shown in blue, posterior samples in gray, and the ground truth in lime. 
   }
         \label{fig:2d-xis}
\end{figure*}

\begin{figure*}[!htbp]
   \centering
   \includegraphics[width=1.\textwidth]{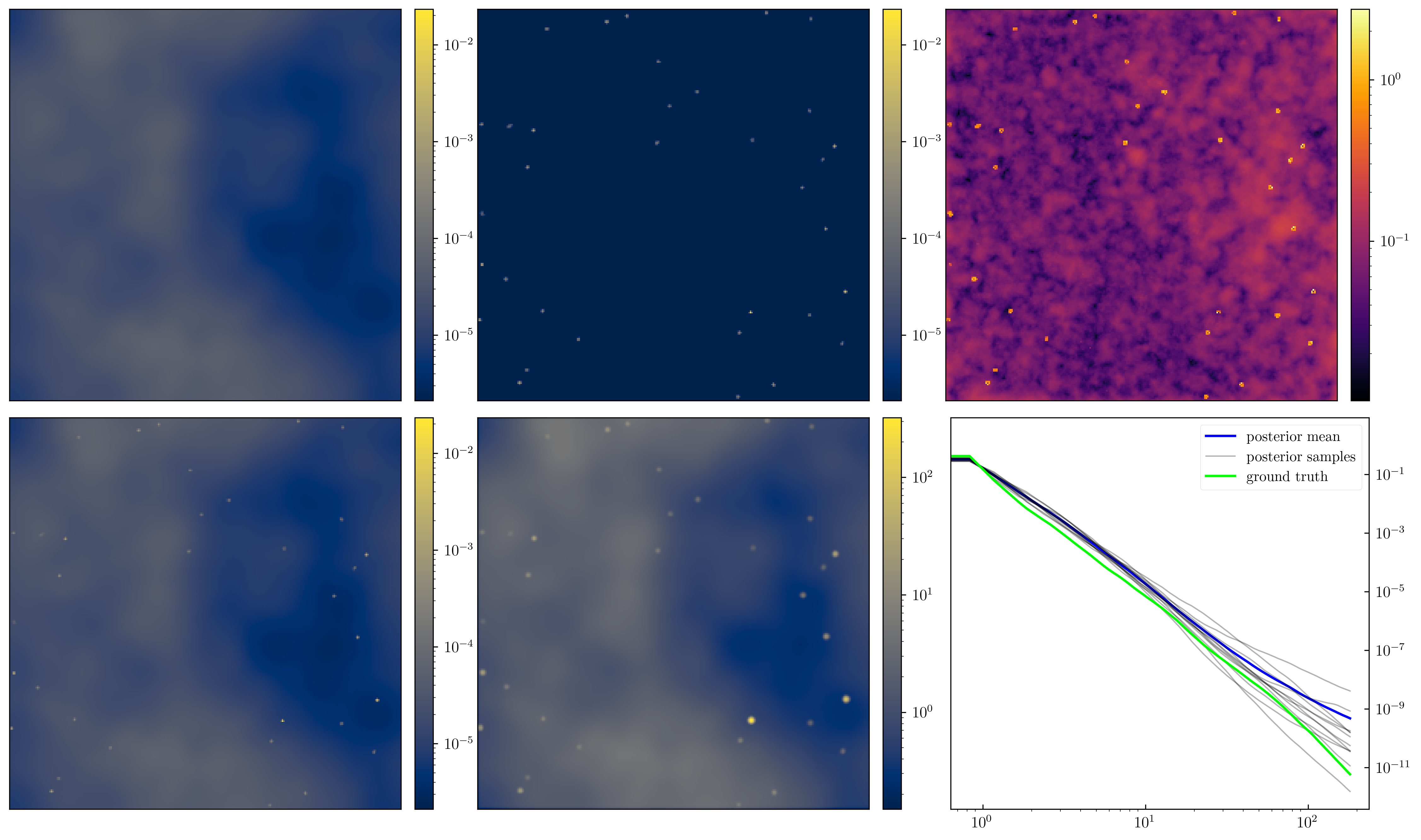}
   \caption{Results of synthetic component separation example for imaging.
   \textit{The top row} shows the reconstructed (posterior mean) diffuse-field emission (left), the reconstructed point-source component (middle), and the relative uncertainty of the sky reconstruction (right), defined as the posterior sample standard deviation divided by the posterior mean.
   Notably, the relative uncertainty is highest at locations containing point sources, as individual posterior samples place them at slightly different positions due to the inferred positional uncertainty.
   \textit{The bottom row} presents the reconstructed sky emission as the sum of the diffuse and point-source components (left), the convolved and exposure-multiplied reconstructed sky emission (middle), and a comparison of the reconstructed spatial power spectrum (posterior mean in blue, posterior samples in gray) with the ground truth power spectrum (in lime, right).
   }
         \label{fig:2d-results}
\end{figure*}

If we now perform a geoVI data fit using our diffuse emission background field $b(\vb{x})$, we obtain a first estimate of the total sky emission field $s(\vb{x})$.
The result of this procedure is shown in Fig.~\ref{fig:2d-xis}.
Similarly to the one-dimensional example, we note that the reconstruction exhibits clear artifacts at the locations in which the diffuse field $b(\vb{x})$ is used to model point-source flux $f(\vb{x})$.
Moreover, since the diffuse field model is forced to explain point-like structures, the inferred spatial power spectrum $\tau_{\vb{x}}$ exhibits excess power at high-frequency modes (see right panel of Fig.~\ref{fig:2d-xis}).
In particular, a pronounced kink around $k \sim 35$ -- corresponding to the typical size of the reconstructed point sources -- drives characteristic artifacts in the background reconstruction, such as the low-flux coronas surrounding the point sources visible in the left panel of Fig.~\ref{fig:2d-xis}.
As a consequence, the reconstructed field (left panel in Fig.~\ref{fig:2d-xis}) will appear rougher than the original signal, since our prior on the diffuse field assumes statistical homogeneity.
By inspecting the latent-space excitations $\xi$ in position space, we can identify locations of model misspecification -- specifically, where a point source should be injected into the model.
In this synthetic reconstruction example, we set a threshold of $\bar{\xi} = 0.65$. At locations where the posterior mean of $\xi$ exceeds $\bar{\xi}$, we inject a point source component modeled as described in Eq.~\eqref{eq:point-source-model}.
We initialize the point source position at the detected location and set its flux to the integral of the diffuse field over one \ac{PSF} length.
This procedure is iteratively repeated to detect additional point sources while simultaneously optimizing their fluxes and positions as free parameters, jointly with the background field.
After each detected point source is introduced, the diffuse latent field $\xi$ is locally reset to standard-normal noise to ensure that overlapping flux is re-attributed in subsequent updates.
At convergence -- i.e., when the \ac{KL} divergence is minimized--we obtain the reconstruction shown in Fig.~\ref{fig:2d-results}.
Examining this reconstruction, we find that the signal is accurately recovered down to the noise level. 
For additional reconstruction metrics, see Appendix~\ref{app:2d-example}.
Moreover, the method successfully disentangles and reconstructs both the diffuse and point-source emission components.
Importantly, the reconstruction model differs from the data-generating inverse-Gamma field. This mismatch illustrates that our method can generalize beyond its prior assumptions, making it robust to moderate model misspecification.
This highlights the robustness of our method in realistic scenarios where model misspecification is inevitable.

\subsection{A multi-frequency sky model}\label{sec:mf_model}
An important aspect of this work is the construction of a sky model that captures structure coherently across both space and frequency.
As more astronomical instruments operate across increasingly broad wavelength ranges, it becomes ever more important to leverage the rich correlations encoded in multi-wavelength and multi-messenger observations. 
Modeling the sky independently at each frequency discards valuable physical information and leads to suboptimal inferences.
Astrophysical processes -- such as synchrotron emission, thermal bremsstrahlung, dust scattering, or line emission -- imprint coherent patterns across both the spatial and spectral domains. 
Properly modeling these correlations allows us to regularize reconstructions, enhance sensitivity, and constrain astrophysical parameters more robustly.
To capture these effects while maintaining computational tractability, we extend our generative modeling framework into the frequency domain. 
In this section, we present a model that jointly describes spatial and spectral correlations in both diffuse, point-source, and extended-source emission, forming a consistent multi-frequency sky prior.

\subsubsection{Diffuse emission}
In Sec.~\ref{sec:1d-example}, we introduced a one-dimensional Gaussian process model for diffuse flux, based on the “correlated field model” developed by \citet{arras_cf}. In their formulation, the correlation structure of the emission field is described via its amplitude spectrum along an axis $i$, which may represent space, time, or energy
\begin{equation*}
	A^{(i)}_{kk'} \coloneqq (2\pi)^{D^{(i)}}\, \delta(k-k') \sqrt{p_{T^{(i)}} (\abs{k})},
\end{equation*}
where $p_{T^{(i)}} (\abs{k})$ is the power spectrum along the $(i)$ direction and $D^{(i)}$ is the dimensionality of the $i$-th space.
The full field’s correlation structure is then written as an outer product over all axes (e.g., spatial $x$ and spectral $\nu$)
\begin{equation*}
	A = \bigotimes_{i\in\qty{x, \nu}}\, A^{(i)}_{kk'}.
\end{equation*}
This formulation is flexible and can describe many classes of diffuse emission. 
However, it implicitly assumes independence between axes, which may limit its ability to capture, for example, spatial variations in energy-dependent flux. 
Such correlations are especially important in X-ray and radio astronomy, where non-thermal emission from \ac{AGN} -- due to synchrotron radiation or inverse Compton scattering -- often follows power-law spectra \citep{rybicki_lightman}.
To better capture these correlations, we introduce alternative priors that incorporate astrophysically motivated spectral and spatial dependencies.
Specifically, we model the sky brightness at location $\vb{x}$ and frequency $\nu$ as
\begin{equation}\label{eq:diffuse_mf}
	I^\text{diff}(\vb{x}, \nu) = I^\text{diff}(\vb{x}, \nu_\text{ref})\, \qty(\frac{\nu}{\nu_\text{ref}})^{\alpha(\vb{x})}\, I_\delta(\vb{x}, \nu),
\end{equation}
where $I(\vb{x}, \nu_\text{ref})$ represents the spatial brightness distribution at reference frequency $\nu_\text{ref}$, and $ \alpha(\vb{x})$ is the spatially varying spectral index. The deviation field $I_\delta(\vb{x}, \nu)$ captures departures from the idealized power-law, due to, e.g., thermal emission, spectral curvature, absorption features, or intrinsic source variability.
In this model, $\alpha(\vb{x})$ varies spatially, while $I_\delta(\vb{x}, \nu)$ is a fully space- and frequency-dependent field. 
The multiplicative nature of $I_\delta$ with both the reference brightness and the power-law components introduces degeneracies.
To break these, we enforce $I_\delta(\vb{x}, \nu_\text{ref}) \equiv \mathds{1}$ and further remove degeneracy with the power-law term by parameterizing
\begin{equation}\label{eq:frequency_deviations}
	I_\delta(\vb{x}, \nu) \coloneqq e^{\tilde{\delta}(\vb{x}, \nu)}\, \qty(\frac{\tilde{\nu}}{\nu_\text{ref}})^{-\bar{\delta}(\vb{x})} ,
\end{equation}
where $\tilde{\delta}(\vb{x}, \nu)$ is modeled as a Gaussian Markov process in the frequency direction -- typically a Wiener process, an integrated Wiener process, or their combination.
The term
\begin{align}\label{eq:slope_removal}
	\bar{\delta}(\vb{x}) \coloneqq 
	\frac{
		\int_{\Omega_\nu} 
		\tilde{\delta}(\vb{x}, \tilde{\nu})\, 
		\log{\frac{\tilde{\nu}}{\nu_\text{ref}}}\, \dd{\tilde{\nu}}
	}{
		\int_{\Omega_\nu} 
		\qty(\log{\frac{\tilde{\nu}}{\nu_\text{ref}}})^2\, \dd{\tilde{\nu}}
	},
\end{align}
where $\Omega_\nu$ is the frequency domain of $I^\text{diff}$, removes the average slope of $\tilde{\delta}$, thereby eliminating degeneracy with the spectral index. (See Appendix~\ref{app:slope_removal} for derivation).
We model the logarithm of the reference frequency brightness distribution $\log{I^\text{diff}(\vb{x}, \nu_\text{ref})}$ and the spectral index $\alpha(\vb{x})$ as correlated fields to learn their spatial and spectral covariance structure. In logarithmic form, the total brightness is expressed as
\begin{equation*}
	\log I^\text{diff}(\vb{x}, \nu) = \log{I^\text{diff}(\vb{x}, \nu_\text{ref})}	
	+ \alpha(\vb{x})\, \log{\qty(\frac{\nu}{\nu_\text{ref}})}
	+ \log I_\delta(\vb{x}, \nu).
\end{equation*}
In this formulation, all components -- spatial brightness, spectral index, and deviations -- are parametrized in harmonic space, while the total brightness field $\log I^\text{diff}(\vb{x}, \nu)$ remains defined in real space and is reconstructed via an inverse Fourier transform
\begin{align}\label{eq:mf_model}
    \log I^\text{diff}(\vb{x}, \nu) &\coloneqq \mathcal{F}_{\vb{x}\vb{k}} \left[
        \frac{\mathcal{A}^\text{spat}_{\vb{k}}}{\int \mathcal{A}^\text{spat}_{\vb{\tilde{k}}} \, \dd{\vb{\tilde{k}}}} \, \sigma^{I_\text{ref}} \, \xi_{\vb{k}}^{I_\text{ref}} \right. \nonumber \\
    &\left. \quad + \frac{\mathcal{A}^\text{spec}_{\vb{k}}}{\int \mathcal{A}^\text{spec}_{\vb{\tilde{k}}} \, \dd{\vb{\tilde{k}}}} \, \qty(\sigma^\text{spec} \, \xi_{\vb{k}}^\text{spec} \, \log{\frac{\nu}{\nu_\text{ref}}} + \delta_{\vb{k}\nu})
    \right]  \\
    &\quad + \alpha_0 \log{\frac{\nu}{\nu_\text{ref}}} + \log{I_{0}}. \nonumber
\end{align}
This construction defines a hierarchical, non-stationary prior over $\log I^\text{diff}(\vb{x}, \nu)$, capturing both spatial and spectral correlations with explicit control over power-law structure and deviations.
Here, $\mathcal{A}^\text{spat}_{\vb{k}}$ and $\mathcal{A}^\text{spec}_{\vb{k}}$ are the amplitude spectra encoding spatial and spectral correlations, respectively; $\xi^\text{spat}_{\vb{k}}$ and $\xi^\text{spec}_{\vb{k}}$ are the corresponding Gaussian field excitations. The parameters $\sigma^{I_\text{ref}}$ and $\sigma^\text{spec}$ set the amplitudes of the fluctuations of the spatial brightness and spectral index. 
Lastly, $\log I_0$ and $\alpha_0$ represent global offsets in brightness and spectral slope.

\subsubsection{Point source emission}
To further extend our prior modeling of the sky across multiple wavelengths, we now incorporate spectral information into our point-source emission model. 
We begin by introducing the point source brightness from Eq.~\eqref{eq:point-source-model} at the reference frequency $\nu_\text{ref}$ as
\begin{equation*}
	I^\text{ps}(\vb{x}, \nu_\text{ref}) = \sum_{i=0}^{N_\text{ps}} I^\text{ps}_i(\vb{x}, \nu_\text{ref}) \coloneqq \sum_{i=0}^{N_\text{ps}} \delta(\vb{x}-\vb{x}_i)\, f_i,
\end{equation*}
where $f_i$ denotes the flux of the $i^\text{th}$ point source, typically drawn from an inverse-Gamma distribution to account for the heavy-tailed nature of astrophysical source populations.
The extension to multiple frequencies is then natural. Drawing inspiration from the diffuse emission model in Eq.~\eqref{eq:diffuse_mf}, we define the point source field across frequency as
\begin{equation}\label{eq:mf_ps}
	I^\text{ps}(\vb{x}, \nu) = \sum_{i=0}^{N_\text{ps}} I^\text{ps}_i(\vb{x}, \nu_\text{ref})\, \qty(\frac{\nu}{\nu_\text{ref}})^{\alpha^\text{ps}_i}\, I^\text{ps}_{\delta,i}(\nu).
\end{equation}
Here, $\alpha^\text{ps}(\vb{x})$ is the spectral index of the $i^\text{th}$ source, and $I^\text{ps}_{\delta,i}(\nu)$ captures deviations from a strict power-law spectrum.
In contrast to the diffuse component, these spectral parameters are assumed to be uncorrelated across sources. 
This independence reflects the physical reality that different point sources often originate from distinct emission mechanisms or environments.
As before, we address the degeneracy between the power-law component and its frequency-dependent deviation field by subtracting the average slope from $\tilde{\delta}_i(\nu)$, using the term $\bar{\delta}$ defined in Eq.~\eqref{eq:slope_removal}. 
This ensures that $\alpha^\text{ps}_i$ remains identifiable even in the presence of broad spectral deviations.

\subsubsection{Extended source emission}
So far we have built up a model to describe the distribution of diffuse and point-source emission across multiple wavelengths.
However, many observed fields contain additional astrophysical structures -- such as radio galaxies, galaxy clusters, or supernova remnants -- that exhibit spatial extents and spectral behaviors distinct from the surrounding background or compact sources.
To accurately capture these components, and to separate them cleanly from diffuse and point-source fields, we introduce a third component: the extended source field.
These sources are neither spatially unresolved nor fully diffuse, and thus require a dedicated prior that reflects their intermediate nature.
Following a structure analogous to the point-source model described above, we represent the extended-source emission as
\begin{equation}
	I^\text{es}(\vb{x}, \nu) = \sum_{i=1}^{N_\text{es}} I^\text{es}_i(\vb{x}, \nu_\text{ref})\, \qty(\frac{\nu}{\nu_\text{ref}})^{\alpha_i^\text{es}(\vb{x})}\, I^\text{es}_{\delta,i}(\vb{x}, \nu),
\end{equation}
where $I^\text{es}_i(\vb{x}, \nu_\text{ref})$ denotes the spatial brightness distribution of the $i$-th extended source at the reference frequency $\nu_\text{ref}$, with $i \in \qty{1, \dots, N_\text{es}}$ and $N_\text{es}$ denoting the number of extended sources in the field.
Unlike point sources, which are modeled as Dirac delta functions, the extended sources are assumed to have nonzero spatial extent.
Specifically, we follow the approach introduced in \citet{lenscharm} for modeling the spatial distributions of strongly lensed galaxies at the reference frequency $\nu_\text{ref}$
\begin{equation}\label{eq:ext_sources}
	I^\text{es}_i(\vb{x}, \nu_\text{ref}) = \gamma_i(\vb{x}, \vb*{\theta}_i)\, I^\text{es}_i(\vb{x}),
\end{equation}
where $\gamma_i(\vb{x}, \vb*{\theta}_i)$ is a parametric envelope function depending on shape parameters $\vb*{\theta}_i$, such as the mean (source location) and covariance (extent and ellipticity), typically instantiated as a Gaussian profile.
The underlying $I^\text{es}_i(\vb{x})$ field is then modeled as a correlated Gaussian process, similar to our diffuse emission model.
The spectral index $\alpha_i^\text{es}(\vb{x})$ governs the power-law scaling of each source's brightness with frequency, and may vary independently from source to source.
Spectral deviations from the pure power-law form -- arising from physical phenomena such as spectral curvature, absorption, or thermal emission -- are described by the term $I^\text{es}_{\delta,i}(\vb{x}, \nu)$, defined analogously to Eq.~\eqref{eq:frequency_deviations}.
As with point sources, we assume the spectral parameters of extended sources are spatially uncorrelated between sources, while allowing for internal spectral correlations across frequency within each individual source.

\subsubsection{Multi-frequency sky model}
Taking all components into consideration, we can now assemble a comprehensive prior model for the sky brightness distribution across multiple wavelengths.
This model remains highly flexible, yet explicitly incorporates spatial and spectral correlations in a physics-informed manner.
Such a construction allows for realistic astrophysical priors while preserving computational tractability, and enables the reconstruction of signals with strong spectral-spatial entanglement -- often present in real multi-frequency observations.

\begin{figure*}
   \centering
   \includegraphics[width=1.\textwidth]{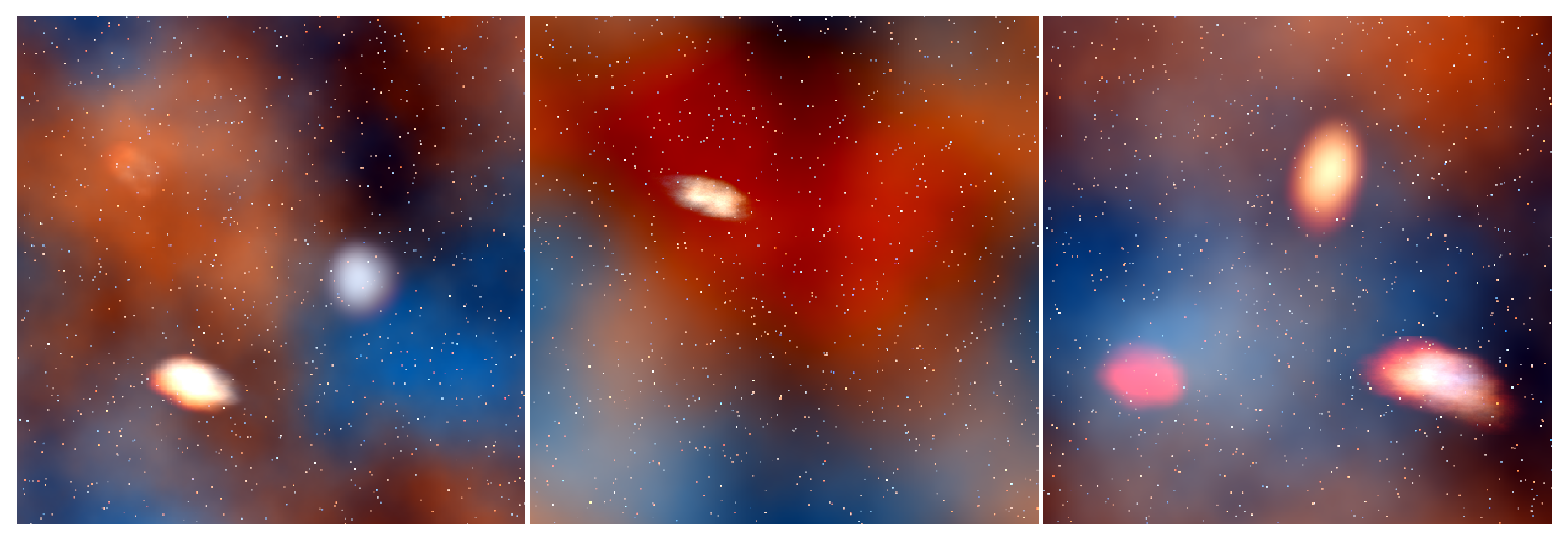}
    \caption{
	Prior samples from the multi-frequency sky model in Eq.~\eqref{eq:mf_sky}.
	Each panel shows one realization of the sky brightness across three frequency channels (columns), with $N_\text{ps} = 1000$ point sources and $N_\text{es} = 3$ extended sources.
	The samples illustrate the interplay of diffuse, point-like, and extended components, all drawn independently from their respective priors.
	Each image is rendered as an RGB composite using three selected frequency bands, with intensity displayed on a logarithmic scale to enhance the visibility of both faint and bright features.
    }
   \label{fig:mf_priors}
\end{figure*}

Given a field of view containing $N_\text{ps}$ point sources and $N_\text{es}$ extended sources, the total sky brightness is modeled as the sum of three emission components
\begin{equation}\label{eq:mf_sky}
	I^\text{sky}(\vb{x}, \nu) = I^\text{diff}(\vb{x}, \nu) + I^\text{ps}(\vb{x}, \nu) + I^\text{es}(\vb{x}, \nu).
\end{equation}
Here, $I^\text{diff}$ captures spatially correlated diffuse emission with spectral coherence, $I^\text{ps}$ accounts for unresolved point sources with independent power-law scaling, and $I^\text{es}$ models spatially extended objects with richer morphologies and spectral structure.
Figure~\ref{fig:mf_priors} shows representative prior samples drawn from this model. 
Visually, they resemble realistic astrophysical fields, capturing the spatial and spectral diversity found in broadband sky surveys.
The model’s flexibility is reflected in the diverse morphologies and spectral behaviors of sources.
For example, in the left panel of Fig.~\ref{fig:mf_priors}, the sources exhibit markedly different spatial and spectral correlation lengths, while in the center panel, two of the extended sources are invisible due to their low brightness relative to the diffuse and point-source background.

\subsection{Synthetic multi-frequency imaging example}\label{sec:mf_example}
Now that we have defined a multi-frequency sky model, we extend the synthetic example from Sec.~\ref{sec:imaging_example} into the spectral domain.
To this end, we use the sky distribution from that example as the reference-frequency brightness distribution in the multi-frequency model of Eq.~\eqref{eq:mf_sky}, and generate a synthetic realization for the associated spectral components.
For simplicity, this example omits extended source emission, i.e., we set $I^\text{es}(\vb{x}, \nu) \equiv 0$.

\begin{figure*}
   \centering
   \includegraphics[width=1.\textwidth]{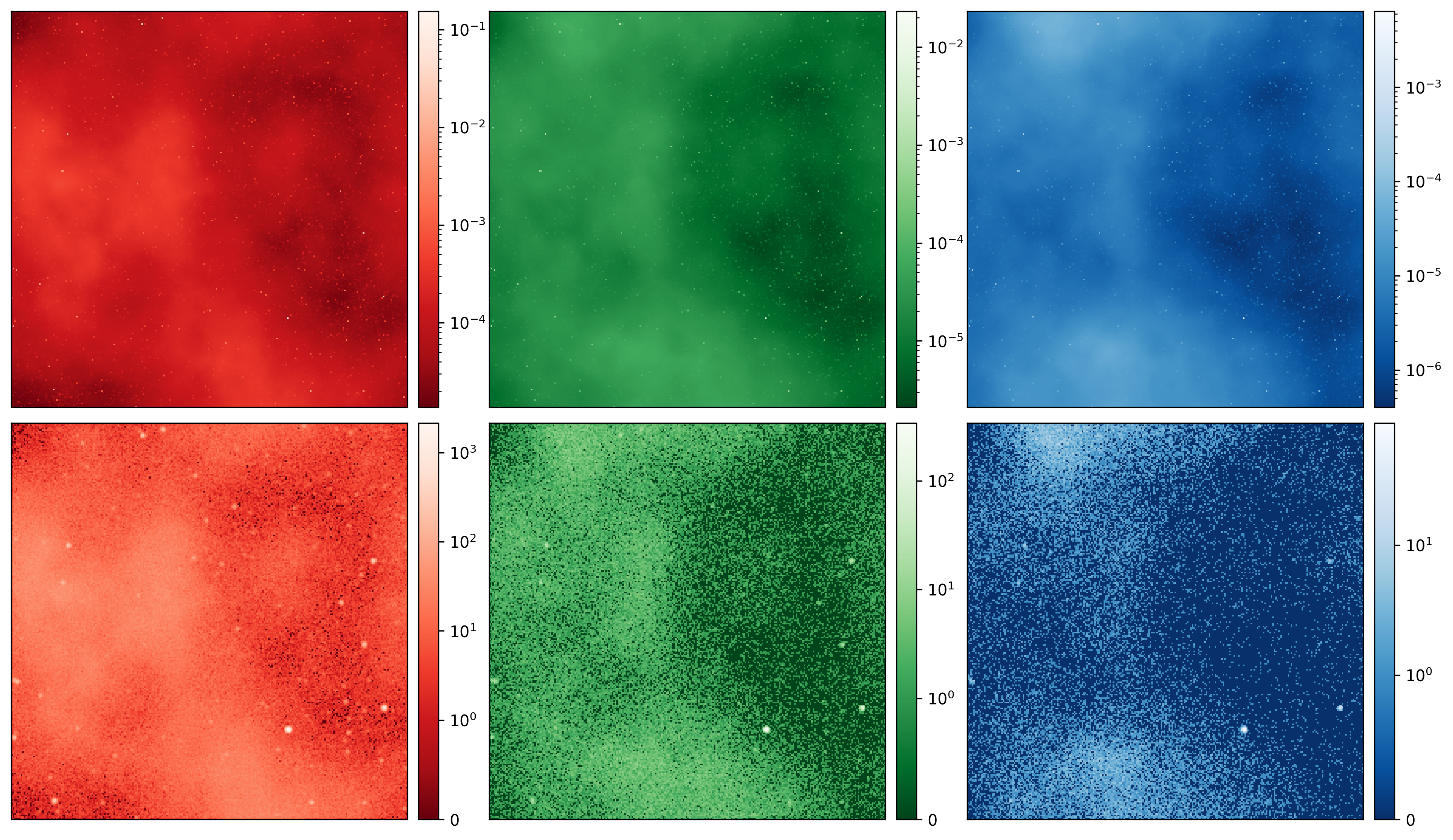}
\caption{
Setup of the synthetic multi-frequency component separation example in imaging.  
\textit{Top row:} synthetic sky emission realizations for each frequency channel -- $\nu_\text{low} = 0.2\text{ – }1.0\,\mathrm{keV}$, $\nu_\text{ref} = 1.0\text{ – }2.0\,\mathrm{keV}$, and $\nu_\text{hi} = 2.0\text{ – }4.5\,\mathrm{keV}$ -- shown in red, green, and blue, respectively.  
\textit{Bottom row:} corresponding synthetic observations after convolution with a Gaussian \ac{PSF} and multiplication by a spatially constant exposure time of \SI{80000}{\second}.
}
   \label{fig:mf_3d_setup}
\end{figure*}

We divide the spectral domain into three non-overlapping bands: 
$\nu_\text{low} = \SIrange{0.2}{1.0}{\kilo\electronvolt}$, 
$\nu_\text{ref} = \SIrange{1.0}{2.0}{\kilo\electronvolt}$, and 
$\nu_\text{hi} = \SIrange{2.0}{4.5}{\kilo\electronvolt}$.
The setup of this inference problem -- including the ground-truth signal and the synthetic data across the three bands -- is shown in Fig.~\ref{fig:mf_3d_setup}.
To reconstruct the sky brightness, we proceed similarly to Sec.~\ref{sec:imaging_example}, now using the full multi-frequency diffuse model $I^\text{diff}(\vb{x}, \nu)$.
For the frequency-dependent deviation field $\tilde{\delta}(\vb{x}, \nu)$ in Eq.~\eqref{eq:frequency_deviations}, we model spectral variability as a Wiener process
\begin{equation*}
	\tilde{\delta}(\vb{x}, \nu) \coloneqq \sigma_\text{dev} \int_{\Omega_\nu} \xi_{\vb{x}, \tilde{\nu}}^\text{dev}\, \dd{\tilde{\nu}},
\end{equation*}
where again $\mathcal{P}(\xi_{\vb{x}, \tilde{\nu}}^\text{dev}) = \mathcal{G}(\xi_{\vb{x}, \tilde{\nu}}^\text{dev}, \mathds{1})$ is a white Gaussian field, and $\sigma_\text{dev}$ steers the amplitude of the spectral deviations.
Compared to the single-frequency example in Sec.~\ref{sec:imaging_example}, this multi-frequency setup provides significantly more constraining power.
In particular, spectral information greatly improves the ability to separate overlapping astrophysical components.
This reflects the physical reality that it is extremely unlikely for two unrelated sources to share identical spectral characteristics.
This additional information is encoded in the richer set of latent fields, including the spatial excitation $\xi_{\vb{x}}^\text{spat}$, the spectral-index excitation $\xi_{\vb{x}}^\text{spec}$, and the frequency deviations excitation field $\xi_{\vb{x}, \nu}^\text{dev}$.

\begin{figure*}
   \centering
   \includegraphics[width=1.\textwidth]{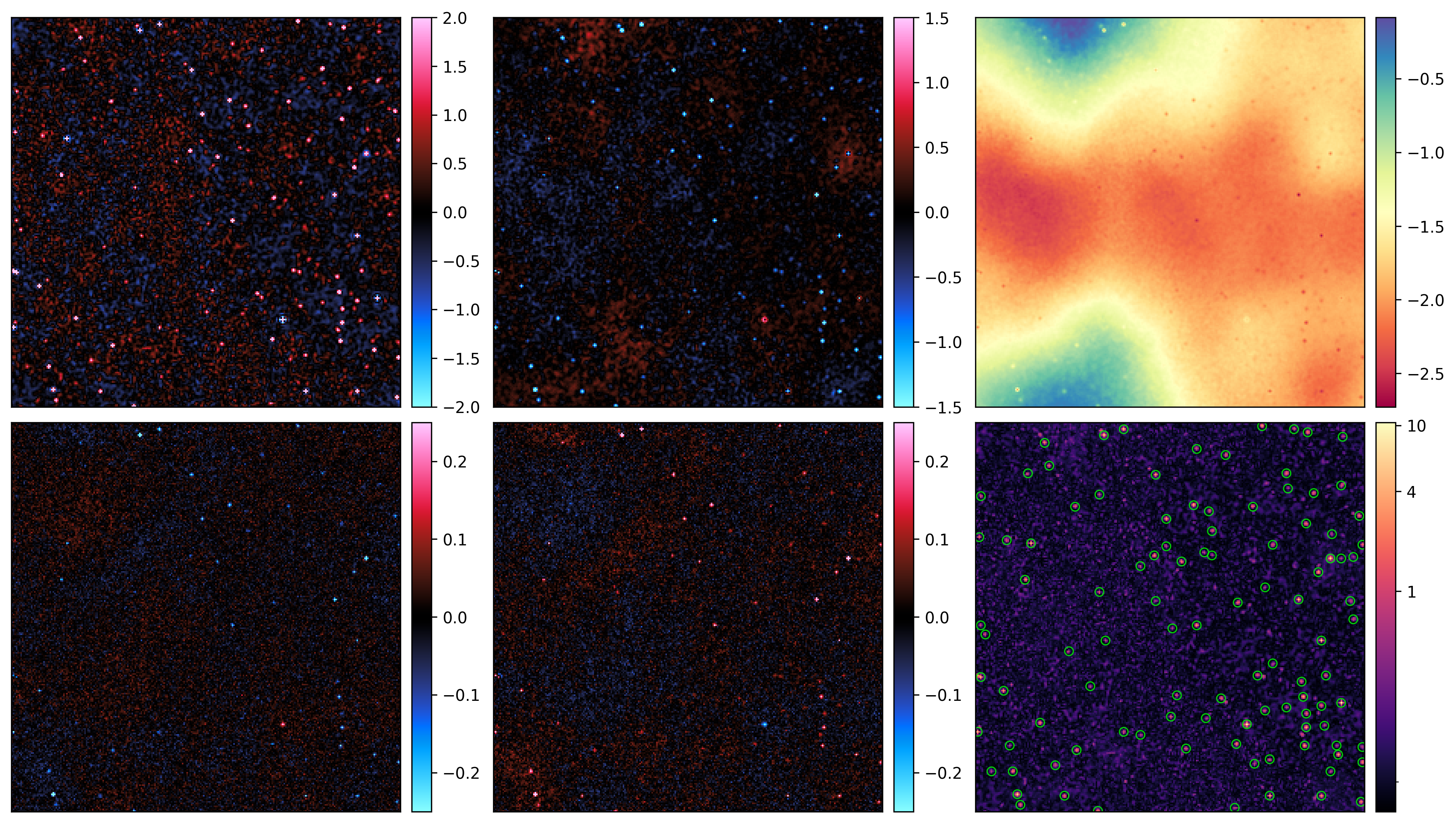}
    \caption{
	Result of diffuse-only reconstruction.
	\textit{Top row:} posterior mean of the spatial excitation field $\xi_{\vb{x}}^\text{spat}$ (left), the spectral-index excitation field $\xi_{\vb{x}}^\text{spec}$ (center), and the inferred spectral index field $\alpha(\vb{x})$ (right) after reconstructing the sky using only the diffuse component.
	\textit{Bottom row:} posterior mean of the spectral deviation excitation field $\xi_{\vb{x}, \nu}^\text{dev}$ at $\nu_\text{low}$ (left), at $\nu_\text{hi}$ (center), and the excitation-norm diagnostic $r(\vb{x})$ (right), computed from the displayed excitation fields.
	Candidate point-source locations identified using the excitation-norm threshold $\bar{r} = 0.6$ are highlighted with lime-colored circles in the $r(\vb{x})$ panel.
}
   \label{fig:mock_xi_3d}
\end{figure*}

To make use of the excitation fields for identifying potential model misspecification -- such as locations where a point-source component may be missing -- we seek a way to combine all latent-space excitations into a single scalar diagnostic.
Specifically, we are interested in locations where some or all of the posterior excitation fields deviate significantly from their prior distribution, as such deviations may indicate unmodeled structure in the data.
We define the excitation-norm diagnostic as
\begin{equation*}
	r(\vb{x}) \coloneqq \sqrt{\qty(\xi_{\vb{x}}^\text{spat})^2 + \qty(\xi_{\vb{x}}^\text{spec})^2 + \sum_{\tilde{\nu}} \qty(\xi_{\vb{x}, \tilde{\nu}}^\text{dev})^2}.
\end{equation*}
Since all excitation fields are a priori white Gaussian fields, the quantity $r(\vb{x})$ follows a Chi distribution with $N_\nu + 1$ degrees of freedom, where $N_\nu$ is the number of frequency channels in the model. 
Posterior deviations from this distribution can be used to flag candidate locations for inserting additional model components, such as unresolved point sources.
When we reconstruct the sky signal shown in Fig.~\ref{fig:mf_3d_setup} using only the diffuse component $I^\text{diff}(\vb{x}, \nu)$, the resulting posterior excitation fields -- together with the excitation-norm diagnostic -- are shown in Fig.~\ref{fig:mock_xi_3d}.

Investigating the reconstructed excitation-norm diagnostic field, we find numerous locations suggestive of model misspecification.
In particular, because the diffuse model is forced to explain point-like features present in the data, the inferred spectral index map $\alpha(\vb{x})$ shows artificially sharp structures which resemble point sources.
To address this, we use the excitation-norm diagnostic $r(\vb{x})$ to identify candidate locations for inserting point-source components.
Specifically, we select all positions where $r(\vb{x}) > \bar{r}$, with a detection threshold of $\bar{r} = 0.6$.
At these positions, we introduce point-source model components $I^\text{ps}(\vb{x}, \nu)$, drawn from the model described in Eq.~\eqref{eq:mf_ps}.

\begin{figure*}
   \centering
   \includegraphics[width=1.\textwidth]{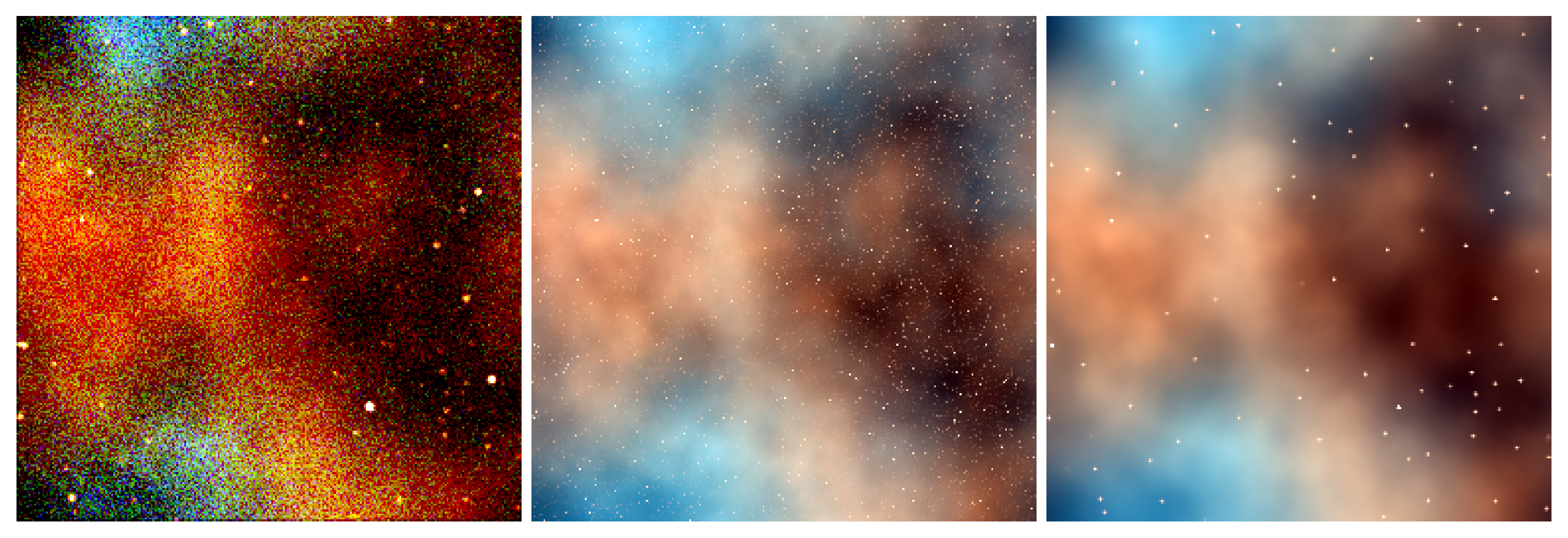}
    \caption{
    Results of the synthetic multi-frequency imaging reconstruction example.
    \textit{Left:} exposure-corrected synthetic data. 
    \textit{Middle:} ground-truth sky emission 
    \textit{Right:} posterior mean of the reconstructed sky.
    All images combine the three energy bands into an RGB composite, where red corresponds to the $\nu_\text{low}$ band, green to $\nu_\text{ref}$, and blue to $\nu_\text{hi}$.
    }
   \label{fig:mock_results_3d}
\end{figure*}

The resulting improved sky reconstruction -- with the model now including both diffuse and point-source components -- is shown in Fig.~\ref{fig:mock_results_3d}.
As expected, the additional information contained in the multi-frequency data improves the reconstruction and allows for more accurate separation into distinct components.
We note that our reconstruction model -- where only explicitly detected point sources are modeled -- differs from the synthetic data-generating model, which assumes a point source at every pixel. Consequently, the reconstructed diffuse background $I^\text{diff}(\vb{x}, \nu)$ must absorb the flux from undetected point sources.
From a statistical perspective, if numerous low-flux sources contribute significantly to the inverse-Gamma distribution’s probability mass, their combined flux may introduce an overall offset in the expected photon rate. This additional flux must then be compensated for by the diffuse background component, potentially leading to its overestimation.
However, this scenario is somewhat artificial: in real observations, the fraction of background flux originating from unresolved sources is typically unknown, as it may also be genuinely diffuse or absorbed by intervening material, such as dust.
For additional diagnostics and a quantitative evaluation of the reconstruction accuracy, we refer to Appendix~\ref{app:3d-example}.

\subsection{LMC SN1987A eROSITA reconstruction}\label{sec:erosita}
We now demonstrate the applicability of our method using real X-ray data from the early data release (EDR) of the SRG/eROSITA observatory.

\subsubsection{Data}
\begin{figure*}[!htbp]
   \centering
   \includegraphics[width=1.\textwidth]{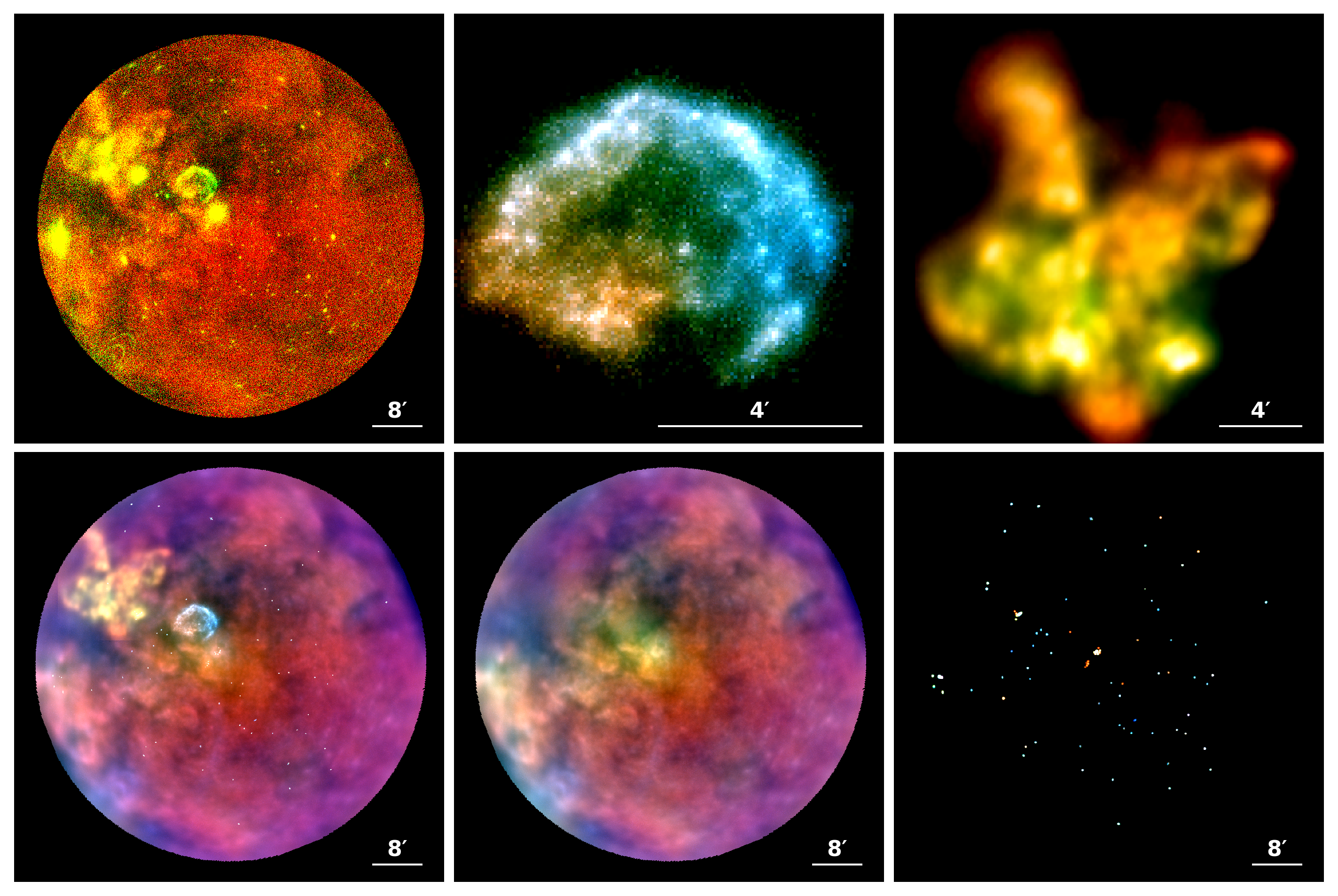}
    \caption{Reconstruction results from the eROSITA early data release (EDR) observation of SN1987A in the LMC.
    Top row (left to right): Exposure- and effective-area-corrected observed data, reconstructed extended emission associated with 30 Doradus Cx, and reconstructed extended emission from the Tarantula Nebula.
    Bottom row (left to right): Reconstructed total sky signal (including all modeled components), reconstructed diffuse background emission (excluding point and extended sources), and reconstructed point-source emission field.
    The point source field has been convolved with a Gaussian kernel with a standard deviation of $\SI{4}{\arcsecond}$.
    All panels employ the same RGB color scheme, representing intensities in the three energy bands: red ($\nu_\text{low}=0.2\text{ - }1.0\,\mathrm{keV}$), green ($\nu_\text{ref}=1.0\text{ - }2.0\,\mathrm{keV}$), and blue ($\nu_\text{hi}=2.0\text{ - }4.5\,\mathrm{keV}$).
    }
   \label{fig:erosita_results}
\end{figure*}

To illustrate the capabilities of our method in a realistic astrophysical scenario, we apply it to an eROSITA pointed observation of SN1987A in the Large Magellanic Cloud (LMC). 
SN1987A, a well-studied supernova remnant, offers an ideal test case because of its complex combination of diffuse emission, compact sources, and extended structures, including shocks, interaction regions, and star-forming regions.
We processed the raw photon-count data using the eROSITA Science Analysis Software System (eSASS), correcting for instrumental exposure, vignetting, and detector characteristics. 
The observations were centered at a celestial position of $\mathrm{RA} = 83.78^\circ$, $\mathrm{DEC} = -69.31^\circ$ and binned into three distinct energy bands following the procedure outlined by \citet{eberle24arxiv}:
$\nu_\text{low} = \SIrange{0.2}{1.0}{\kilo\electronvolt}$,
$\nu_\text{ref} = \SIrange{1.0}{2.0}{\kilo\electronvolt}$, and
$\nu_\text{hi} = \SIrange{2.0}{4.5}{\kilo\electronvolt}$.
For the present analysis, we utilized data from eROSITA telescope modules (TMs) TM$1$, TM$3$, and TM$4$.
This selection was motivated by computational efficiency: each TM has a distinct instrumental response, so including all five available TMs would significantly increase computational costs.
While using more TMs would improve the signal-to-noise ratio and could potentially enhance reconstruction accuracy and point-source detectability, we found that restricting the analysis to three modules offers a favorable trade-off between fidelity and runtime.
Moreover, the primary objective of this application is not to achieve maximum reconstruction fidelity, but rather to showcase the latent-space tension diagnostic on real observational data.
The resulting dataset, shown in the top-left panel of Fig.~\ref{fig:erosita_results}, provides the input to our inference.

\subsubsection{Reconstruction}
\begin{figure}[!htbp]
   \centering
   \includegraphics[width=.5\textwidth]{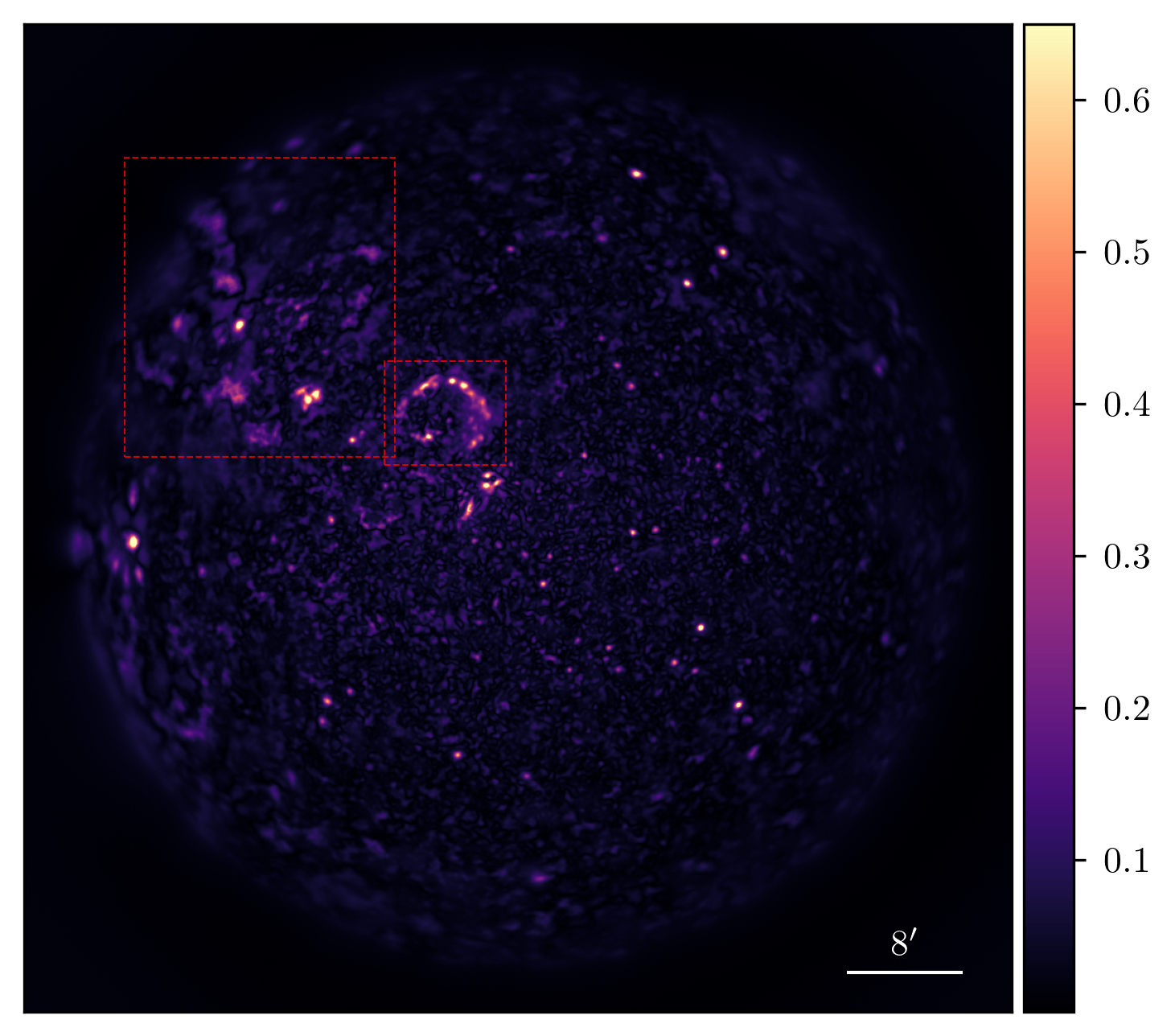}
      \caption{
	Excitation-norm diagnostic map highlighting regions of potential model misspecification after the initial diffuse-only reconstruction of the SN1987A eROSITA dataset. 
	Bright areas represent elevated latent-space excitations, clearly identifying well-known extended sources such as the Tarantula Nebula and 30 Doradus C (highlighted with red boxes). 
	These features motivate the explicit modeling of extended-source components.
        }
         \label{fig:erosita_ext_sources}
\end{figure}

We initially reconstructed the sky emission using our multi-frequency diffuse emission model (Eq.~\eqref{eq:diffuse_mf}).
We implemented the multi-frequency emission model and the eROSITA response using the \texttt{J-UBIK} package \citep{jubik}. 
This first step identifies regions where the diffuse-only assumption is inadequate. 
To diagnose potential model misspecifications, we compute the excitation-norm diagnostic, $r(\vb{x})$, derived from the latent-space excitations. 
Elevated values in this diagnostic map, as shown in Fig.~\ref{fig:erosita_ext_sources}, highlight discrepancies between the assumed smooth, diffuse background emission model and actual astrophysical structures in the observed data.

The excitation-norm diagnostic clearly identifies two prominent extended X-ray sources: the Tarantula Nebula and 30 Doradus C. 
The accurate recovery of these well-known astrophysical features confirms the effectiveness of the latent-space diagnostic as a robust indicator for refining the emission model. 
Consequently, we enhance our reconstruction by explicitly incorporating two extended-source model components at these identified positions, modeled according to Eq.~\eqref{eq:ext_sources}. 
Currently, extended sources are identified manually. 
Automating this step would reduce subjective bias, improve reproducibility, and streamline the analysis pipeline. 
Leveraging neural-network segmentation, \citet{fuchs_in_prep} have achieved good extended-source component separation on radio data. 
Future work might include supplementing their methods with latent-space information.
We note that explicitly modeling these extended-source components substantially reduces residual latent-space excitations, thereby enhancing the physical realism and interpretability of the reconstruction.

To speed up the inference and source detection processes, we initiate our final reconstruction incorporating the two extended source models by reconstructing the sky only at the reference frequency $\nu_\text{ref}$, i.e., $I(\vb{x}, \nu_\text{ref}) = I^\text{diff}(\vb{x}, \nu_\text{ref}) + I^\text{es}(\vb{x}, \nu_\text{ref})$. We utilize latent-space tension at the reference frequency to iteratively introduce reference-frequency point sources, as described in Sec.\ref{sec:imaging_example}. Upon convergence of this reconstruction, we perform a multi-frequency reconstruction using the full model described by Eq.~\ref{eq:mf_sky}.
Due to the clean factorization of our prior into a spatial component defined at the reference frequency and a spatio-spectral component, no intermediate transition step (such as the transition model implemented by \citealt{wmarg}) is required, significantly accelerating the reconstruction process. The inference proceeds iteratively, simultaneously optimizing for diffuse, point-source, and extended-source components until convergence. 
Point sources are introduced wherever significant features appear in the excitation-norm diagnostic map, as detailed in Sec.~\ref{sec:mf_example}. To enhance the efficiency and accuracy of point-source detection, we apply a matched filter to the excitation-norm diagnostic map (see Appendix~\ref{app:ps_filtering} for details).
As detection criteria, we incorporate a point-like source into our forward model at locations where the excitation-norm diagnostic exceeds a chosen threshold ($0.6$ for unfiltered and $0.3$ for filtered diagnostics). Convergence is defined as stabilization of the Kullback-Leibler divergence difference between consecutive iterations, accompanied by the absence of significant features in the excitation-norm diagnostic map over ten consecutive iterations.

We present the results of applying this method in the following section.

\subsubsection{Results}
The main results from our multi-component reconstruction of the eROSITA early data release (EDR) observation of SN1987A in the LMC are presented in Fig.~\ref{fig:erosita_results}. 
All astrophysical source components are clearly separated, and no residual point-like structures remain visible within the reconstructed diffuse background.
Given that we model the sky emission as a spatio-spectral field, we can extract the model prediction for the sky (or any component's) emission at any given wavelength. 
We show an example of the predicted sky photon flux at $\SI{2.4}{\kilo\electronvolt}$ in Fig.~\ref{fig:erosita_intermediate_wavelength} in Appendix~\ref{app:erosita_diagnostics}.
Thanks to the explicit modeling of the extended sources 30 Doradus C and the Tarantula Nebula, we are able to recover their spatial and spectral correlation structures from eROSITA data with unprecedented detail.
For a direct comparison with previous IFT-based reconstructions and Chandra observations covering the same region, we refer the reader to \citet{eberle24arxiv}.
Our approach enables subpixel localization of point-like structures and improves over earlier inverse-Gamma-based models by explicitly modeling source positions and allowing for cleaner disentanglement of components. 
A complete list of the detected point-like sources, along with additional posterior diagnostics -- including noise-weighted residual analysis -- is provided in Appendix~\ref{app:erosita_diagnostics}.
Lastly, we emphasize that our Bayesian inference framework not only yields a single reconstruction of the modeled signal fields but also quantifies uncertainties through posterior sampling. 
These posterior samples comprehensively characterize the inferred probability distribution of the reconstructed signals.


%
%
\section{Conclusions}\label{sec:conclusions}
In this work, we presented a novel method for astrophysical component detection and separation in single- and multi-band imaging data. 
Our approach combines a new multi-frequency sky model with a latent-space field-tension diagnostic to automatically identify and disentangle point-like and extended emission from diffuse background structures.
The model was validated on synthetic data, demonstrating accurate component separation and denoising.
In addition to imaging applications, a one-dimensional variant of the method can be used, for instance, for line detection in noisy spectra.

We further applied the method to SRG/eROSITA X-ray observations of SN1987A in the LMC, successfully reconstructing diffuse, extended, and point-source components potentially with sub-pixel accuracy. 
Extended sources like the Tarantula Nebula and 30 Doradus C were explicitly modeled, enabling detailed spatial and spectral reconstruction.
By jointly analyzing multi-frequency data, we recover spatially resolved spectral features and improve physical interpretability. 
Since the sky emission model is fully decoupled from the instrument response, the method is applicable to any telescope or wavelength band, making it well suited for future multi-messenger analyses. 

Looking ahead, IFT-based radio imaging methods such as \texttt{resolve} and \texttt{fast resolve} could benefit from the automatic point-source modeling introduced here.
The Bayesian formulation naturally provides uncertainty quantification through posterior samples, enabling robust flux estimation and residual diagnostics.
The multi-frequency sky model and latent-space detection tools will be made publicly available in \texttt{J-UBIK}\footnote{\url{https://github.com/NIFTy-PPL/J-UBIK}}.

\subsection{Future Work}

Several enhancements to our method are currently under development to extend its performance and applicability. 
A key priority is calibrating point-source detection to achieve catalog-grade completeness and purity, validated through simulations. 
We plan to refine detection diagnostics, incorporating matched-filter techniques to improve sensitivity, particularly for faint and crowded fields.

Automating extended-source identification is another important goal. 
Leveraging latent-space diagnostics, we will develop criteria for algorithmically detecting and modeling extended astrophysical sources, removing manual intervention.

Instrumental modeling improvements, including continuous spatial interpolation of the PSF for the point sources, will be integrated into the forward model, ensuring more precise reconstructions. 
Additionally, we plan to explore broader astrophysical applications, such as cosmic microwave background source detection, spectral-line identification in spectroscopic data, exoplanet transit searches in multi-messenger data, and more.
These ongoing developments aim to provide a robust, versatile, and fully automated Bayesian imaging tool, broadly applicable across astronomical datasets.


\begin{acknowledgements}
M.G., V.E., and M.W. acknowledge financial support from the German Aerospace Center and Federal Ministry of Education and Research through the project \textit{Universal Bayesian Imaging Kit -- Information Field Theory for Space Instrumentation} (F{\"o}rderkennzeichen 50OO2103).
M.G. acknowledges support from the European Union-funded project \texttt{mw-atlas} under grant agreement No.~101166905.
V.E. and P.F. acknowledge funding through the German Federal Ministry of Education and Research for the project ErUM-IFT: Informationsfeldtheorie für Experimente an Großforschungsanlagen (Förderkennzeichen: 05D23EO1).

This work makes use of data from eROSITA, the soft X-ray instrument aboard SRG, a joint Russian-German science mission supported by the Russian Space Agency (Roskosmos), 
in the interests of the Russian Academy of Sciences represented by its Space Research Institute (IKI), and the Deutsches Zentrum für Luft- und Raumfahrt (DLR). The SRG spacecraft 
was built by Lavochkin Association (NPOL) and its subcontractors, and is operated by NPOL with support from the Max Planck Institute for Extraterrestrial Physics (MPE). 
The development and construction of the eROSITA X-ray instrument was led by MPE, with contributions from the Dr. Karl Remeis Observatory Bamberg \& ECAP (FAU Erlangen-Nuernberg), 
the University of Hamburg Observatory, the Leibniz Institute for Astrophysics Potsdam (AIP), and the Institute for Astronomy and Astrophysics of the University of Tübingen, with the support of DLR and the Max Planck Society. 
The Argelander Institute for Astronomy of the University of Bonn and the Ludwig Maximilians Universität Munich also participated in the science preparation for eROSITA.

The eROSITA data shown here was processed using the eSASS software system developed by the German eROSITA consortium.

M.G. thanks Jakob Roth for valuable discussions.
\end{acknowledgements}

%
   \bibliographystyle{aa} 
   \bibliography{bib.bib} 
%


%
%
%

\begin{appendix} 
\section{Latent-space point source representation}\label{app:1d_ps_theory}
In this appendix, we provide a deeper theoretical understanding of how point sources, when incorrectly modeled by a purely diffuse sky component, manifest in the latent-space excitations. 
This insight is crucial for interpreting the excitation maps used in our component-separation method.

\subsection{Gaussian correlated field}
Consider a one-dimensional sky field $s$ composed of a diffuse component and $N_\text{ps}$ point sources
\begin{equation*}
	s^\text{gt}_x \coloneqq \mathcal{F}_{xk} \mathcal{A}^\text{gt}_k \xi^\text{gt}_{k} + \sum_{i=1}^{N_\text{ps}} b_i\, \delta(x - x_i),
\end{equation*}
where we define the fixed amplitude spectrum as $\mathcal{A}_k \coloneqq \sqrt{\hat{T}_k}$.
Suppose we attempt to fit this sky using only a diffuse component
\begin{equation*}
	s_x \coloneqq \mathcal{F}_{xk} \mathcal{A}_k \xi_{k}.
\end{equation*}
We then aim to minimize the functional
\begin{equation}\label{eq:app_loss}
	\mathcal{L} \coloneqq \int_{\Omega_x} \qty(s^\text{gt}_{\tilde{x}} - s_{\tilde{x}})^2 \dd{\tilde{x}} + \int_{\Omega_x} \xi^2_{\tilde{x}} \dd{\tilde{x}},
\end{equation}
where minimizing $\mathcal{L}$ ensures both data fidelity (matching observations) and regularization (enforcing the standard-normal prior on $\xi_x$).
Invoking Plancherel's theorem we can rewrite the integral in Fourier space
\begin{equation*}
\begin{aligned}
	\mathcal{L} &= \frac{1}{(2\pi)^D} \left[\int_{\Omega_k} \mathcal{F}_{\tilde{k}\tilde{x}} (s^\text{gt}_{\tilde{x}} - s_{\tilde{x}})^2 \, \dd{\tilde{k}} + \int_{\Omega_k} \mathcal{F}_{\tilde{k}\tilde{x}}\, \xi^2_{\tilde{x}} \, \dd{\tilde{k}} \right] \\
	&= \frac{1}{(2\pi)^D} \left[\int_{\Omega_k} \mathcal{F}_{\tilde{k}\tilde{x}} \left(
	\mathcal{F}_{\tilde{x}\tilde{k}} \mathcal{A}^\text{gt}_{\tilde{k}}\xi^\text{gt}_{\tilde{k}} + 
	\sum_{i=1}^{N_\text{ps}} b_i\, \delta(\tilde{x} - x_i) - \mathcal{F}_{\tilde{x}\tilde{k}} \mathcal{A}_{\tilde{k}} \xi_{\tilde{k}}
	\right)^2\, \dd{\tilde{k}} \right. \\
	&\quad + \left. \int_{\Omega_k} \mathcal{F}_{\tilde{k}\tilde{x}}\, \xi^2_{\tilde{x}} \, \dd{\tilde{k}} \right].
\end{aligned}
\end{equation*}
This loss is minimized for
\begin{equation*}
\begin{aligned}
 0 &\stackrel{!}{=} \fdv{\mathcal{L}}{\xi'} \\
   &= \frac{2}{(2\pi)^D} \int_{\Omega_k} \delta\xi'_{\tilde{k}}\,
      \Bigl[
        -\mathcal{A}^\text{gt}_{\tilde{k}}\,
         \mathcal{A}_{\tilde{k}}\,\xi^\text{gt}_{\tilde{k}}
        - \mathcal{A}_{\tilde{k}}
          \sum_{i=1}^{N_\text{ps}} 
            b_i\,\mathcal{F}_{\tilde{k}\tilde{x}}\,
            \delta(\tilde{x}-x_i)
        \\[1ex]
        & \quad + \bigl(\mathcal{A}_{\tilde{k}}^2 + 1\bigr)\,\xi_{\tilde{k}}
      \Bigr]
      \,\dd{\tilde{k}}
\end{aligned}
\end{equation*}
which, to hold for all $\delta\xi'$, implies
\begin{equation}\label{eq:linear_xi}
	\xi_{\tilde{k}} = \frac{\mathcal{A}_{\tilde{k}}\mathcal{A}^\text{gt}_{\tilde{k}}\xi^\text{gt}_{\tilde{k}}}{\mathcal{A}^2_{\tilde{k}} + 1} + \frac{\mathcal{A}_{\tilde{k}}}{\mathcal{A}^2_{\tilde{k}} + 1}\, \sum_{i=1}^{N_\text{ps}} b_i\, \mathcal{F}_{\tilde{k}\tilde{x}}\, \delta(\tilde{x} - x_i).
\end{equation}
The first term in Eq.~\eqref{eq:linear_xi} directly relates the inferred excitations to the ground truth excitations scaled by the amplitude spectrum. 
Specifically, if a given Fourier mode k has a high amplitude $\mathcal{A}_k$, the inferred excitation $\xi_k$ is closely aligned with the ground truth excitation $\xi_k^{\text{gt}}$, since the likelihood-like term (first term in the sum in Eq.~\eqref{eq:app_loss}) dominates over the prior-like one (second term in Eq.~\eqref{eq:app_loss}). 
Conversely, if the amplitude is low, the prior enforcing small excitations dominates, thereby pushing the inferred excitation closer to zero and diminishing the influence of the ground truth excitation.
The second term explicitly represents the latent-space excitations induced by point sources. Each point source creates excitations that appear as a superposition of monochromatic waves, whose frequencies are determined by the positions of the point sources in position-space. Crucially, these waves are scaled by the amplitude spectrum $\mathcal{A}_k$, implying that the latent-space signature of a point source is modulated by the correlation structure assumed in the diffuse field model.
Understanding the meaning of the latent-space excitations created by a point source is not just an interesting mathematical exercise.
In fact, convolving the real-space excitations $\xi_x$ with the inverse Fourier transform of the second term template enhances point-source visibility.
\subsection{Lognormal correlated field}
In the case, like the ones analyzed in this work, in which we want to model a lognormal field with point sources, the situation becomes more complicated.
The ground truth sky becomes
\begin{equation*}
	s^\text{gt}_x \coloneqq e^{\mathcal{F}_{xk} \mathcal{A}^\text{gt}_k \xi^\text{gt}_{k}} + \sum_{i=1}^{N_\text{ps}} b_i\, \delta(x - x_i),
\end{equation*}
while the reconstruction sky is
\begin{equation*}
	s_x \coloneqq e^{\mathcal{F}_{xk} \mathcal{A}_k \xi_{k}}.
\end{equation*}
In fact the functional that needs to be minimized in this case becomes
\begin{equation*}
	\mathcal{L} = \int_{\Omega_x} \qty[\qty(e^{\mathcal{F}_{\tilde{x}k} \mathcal{A}^\text{gt}_k \xi^\text{gt}_{k}} + \sum_{i=1}^{N_\text{ps}} b_i\, \delta(\tilde{x} - x_i) - e^{\mathcal{F}_{\tilde{x}k} \mathcal{A}_k \xi_{k}})^2 + \xi^2_{\tilde{x}}]  \dd{\tilde{x}}.
\end{equation*}
This requires solving a system of trascendental equations. We defer the analysis of this system to future work.
In the linear approximation, we can use the point source filter derived in the Gaussian case.

\subsection{Latent-space point source filtering}\label{app:ps_filtering}
Motivated by the theoretical considerations presented above, we now develop a filtering approach for latent-space excitations induced by point sources. 
Filtering latent-space excitations improves sensitivity to faint point sources and significantly reduces false-positive detections, making our approach particularly effective in fields with multiple different source components.

In the Gaussian field case, as discussed previously, the optimal latent-space filter is given by the inverse Fourier transform of the point-source-induced excitation pattern (second term in Eq.~\eqref{eq:linear_xi}), modulated by the amplitude spectrum
\begin{equation*}
	f^\text{lin}_x \coloneqq \mathcal{F}_{x\tilde{k}} \qty[\frac{\mathcal{A}_{\tilde{k}}}{\mathcal{A}^2_{\tilde{k}} + 1}\, \mathcal{F}_{\tilde{k}\tilde{x}} \delta(\tilde{x} - x_0)],
\end{equation*}
where $x_0$ is the center of the filter.
where the filter center is positioned at $x_0$. Here, the delta function in real space transforms into a complex exponential in Fourier space, clearly showing that the filter aligns with monochromatic waves modulated by the amplitude spectrum.
On a grid, the specific shape of the filter then depends on the choice of discrete harmonic transform used.

In the lognormal scenario, it is not straightforward to analytically determine the latent-space response to a point source due to the complexity of the lognormal correlation structure. 
Moreover, the exact latent-space response depends on the accuracy of the posterior approximation -- specifically, how effectively the KL divergence has been minimized at the current iteration.

To overcome these complexities, we empirically construct a heuristic matched filter based on the latent-space response observed in synthetic simulations, such as those presented in Sec.~\ref{sec:mf_example}. 
We define this heuristic filter as
\begin{equation*}
	f^\text{heur}_{\vb{x}} \coloneqq 
	\begin{cases}
		1 & \text{if } |\vb{x}| \leq r_\text{ps}, \\
		-1 & \text{if } r_\text{ps} < |\vb{x}| \leq r_\text{max}, \\
		0 & \text{otherwise}.
	\end{cases}
\end{equation*}
This filter assigns a positive weight within a radius $r_\text{ps}$ around the source center, capturing the primary excitation region, and a negative weight in the surrounding annulus up to a radius $r_\text{max}$, accounting for typical overshoot patterns observed in the latent-space excitation field as seen, e.g., in Fig.~\ref{fig:2d-xis} and Fig.~\ref{fig:mock_xi_3d}.
Exploring rigorous analytical treatments or data-driven methods (e.g., neural-network-based latent-space filters) represents promising future directions to further refine point-source filtering for lognormal correlated fields.

\section{Diagnostics} 
We provide two reconstruction metrics to quantitatively asses the fidelity of the reconstructions presented throughout this work.
We define the posterior mean of the noise weighted residuals of a signal field $s$ as
\begin{equation}\label{eq:nwr}
	\text{nwr} \coloneqq \expval{\frac{R(s) - d}{\sqrt{R(s)}}}_{\mathcal{P}(s|d)} \simeq \expval{\frac{R(s) - d}{\sqrt{R(s)}}}_{\mathcal{Q}(s|d)},
\end{equation}
where the expectation value is evaluated as an average over approximate geoVI posterior $\mathcal{Q}(s|d)$ samples.
Furthermore, for synthetic data simulations in which the ground truth signal is known, we can define the uncertainty weighted residuals as
\begin{equation}\label{eq:uwr}
	\text{uwr} \coloneqq \frac{\expval{s}_{\mathcal{P}(s|d)} - s_\text{gt}}{\sqrt{\expval{\qty(s - \expval{s}_{\mathcal{P}(s|d)})^2}_{\mathcal{P}(s|d)}}} \simeq \frac{\expval{s}_{\mathcal{Q}(s|d)} - s_\text{gt}}{\sqrt{\expval{\qty(s - \expval{s}_{\mathcal{Q}(s|d)})^2}_{\mathcal{Q}(s|d)}}},
\end{equation}
where $s_\text{gt}$ is the ground truth signal. and again the same holds for the expectation values over the approximate posterior $\mathcal{Q}(s|d)$.

\subsection{Imaging example}\label{app:2d-example}
\begin{figure*}[!htbp]
   \centering
   \includegraphics[width=1.\textwidth]{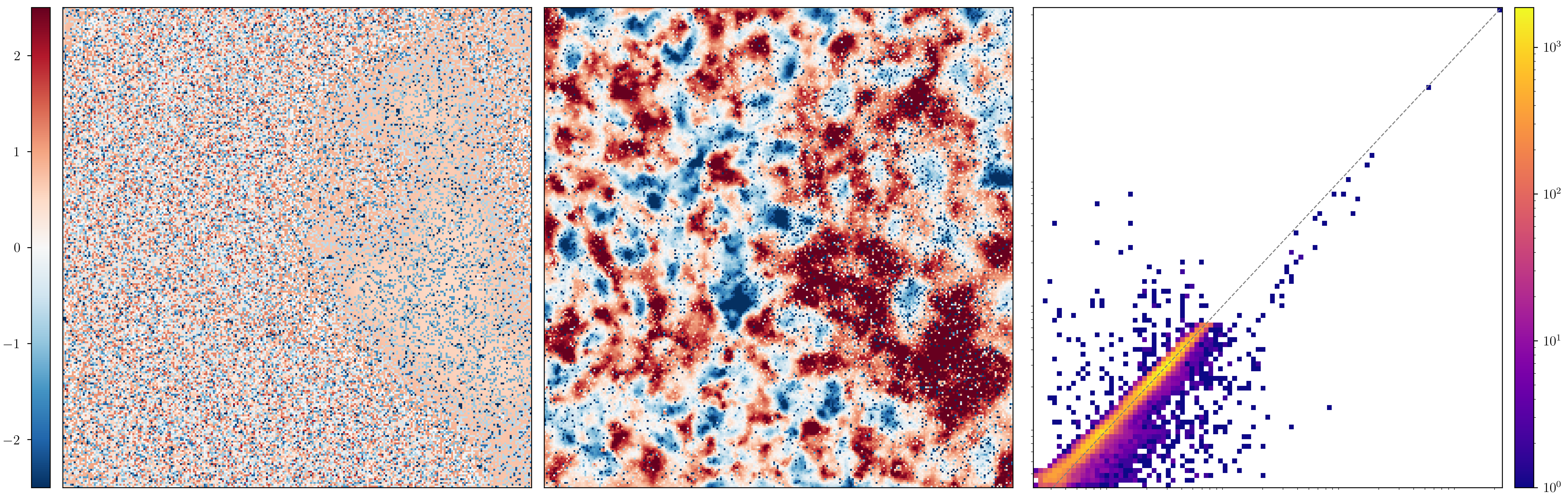}
    \caption{
    Imaging example residuals (for the synthetic data example shown in Fig.~\ref{fig:2d-setup}). 
    \textit{Left panel:} noise-weighted residuals (nwr); \textit{middle panel:} uncertainty-weighted residuals (uwr); \textit{right panel:} reconstructed flux vs. ground-truth flux comparison.
    The nwrs and uwrs share the same color bar.
    }
   \label{fig:2d_demo_residuals}
\end{figure*}

Figure~\ref{fig:2d_demo_residuals} shows both residual metrics for the single-frequency imaging example, showing good reconstruction fidelity.

\subsection{Synthetic multi-frequency imaging example}\label{app:3d-example}
\begin{figure*}[!htbp]
   \centering
   \includegraphics[width=1.\textwidth]{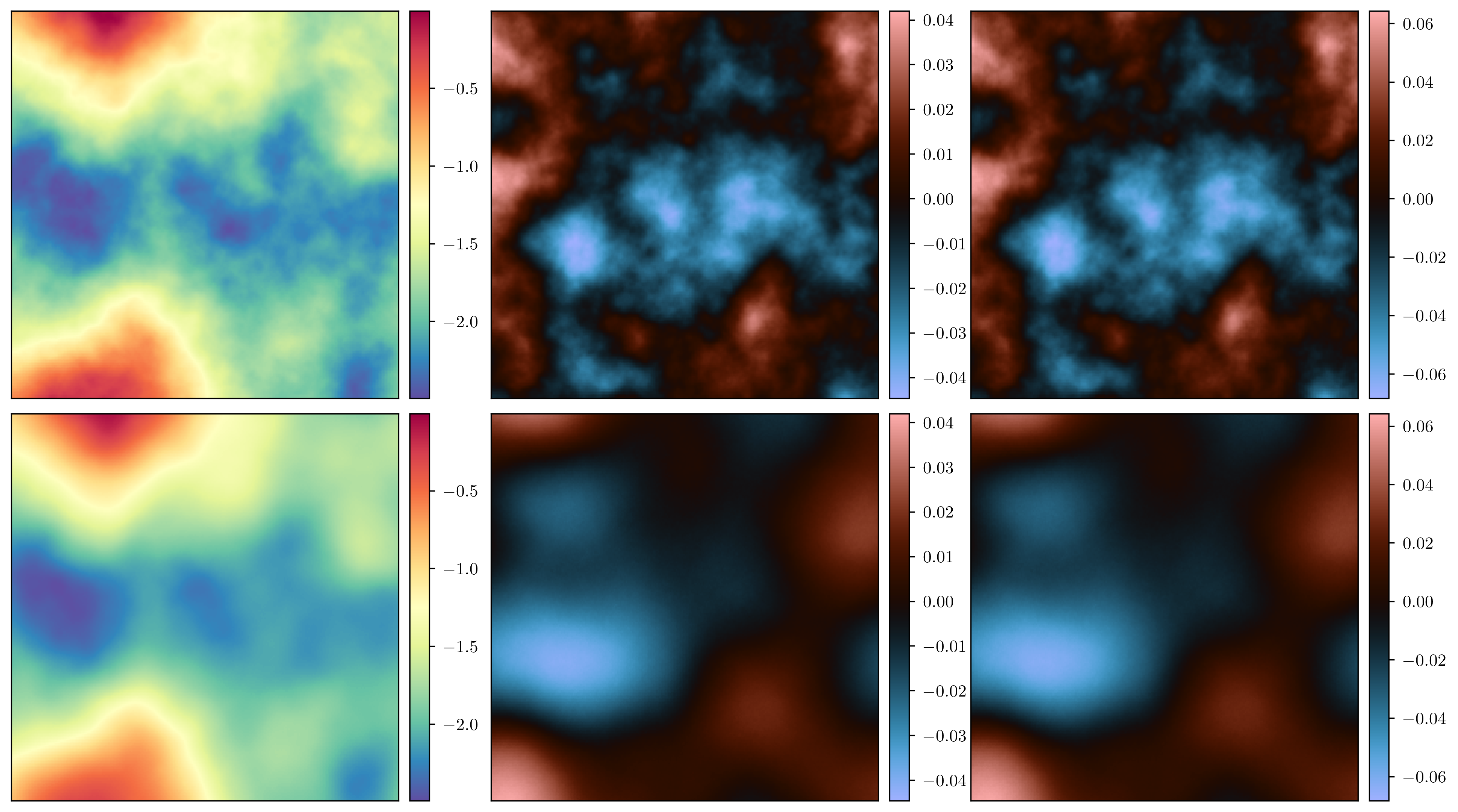}
    \caption{
	Spectral reconstruction from the synthetic multi-frequency imaging example from Sec.~\ref{sec:mf_example}.
	The top row shows the ground truth spectral index map $\alpha(\vb{x})$ (left panels) and spectral deviations $\log I_\delta(\vb{x}, \nu)$ as defined in Eq.~\eqref{eq:diffuse_mf}.
   The central panels represent the spectral fluctuations at $\nu_\text{low}$ and the rigth ones at $\nu_\text{hi}$.
	The bottom row presents the corresponding posterior mean reconstructions. These highlight the method’s ability to recover both large-scale spectral slopes and localized deviations across the frequency range.
    }
   \label{fig:mock_spectral_3d}
\end{figure*}

In Fig.~\ref{fig:mock_spectral_3d}, we present the ground truth and reconstructed spectral index maps and spectral deviation fields, demonstrating that our method accurately captures both large-scale and small-scale spectral features. 
The reconstructed spectral index closely matches the ground truth, confirming that the method reliably separates spectral variations from the primary power-law trend. Similarly, the spectral deviations reconstruction effectively captures localized deviations from the power-law assumption.

\begin{figure*}[!htbp]
   \centering
   \includegraphics[width=1.\textwidth]{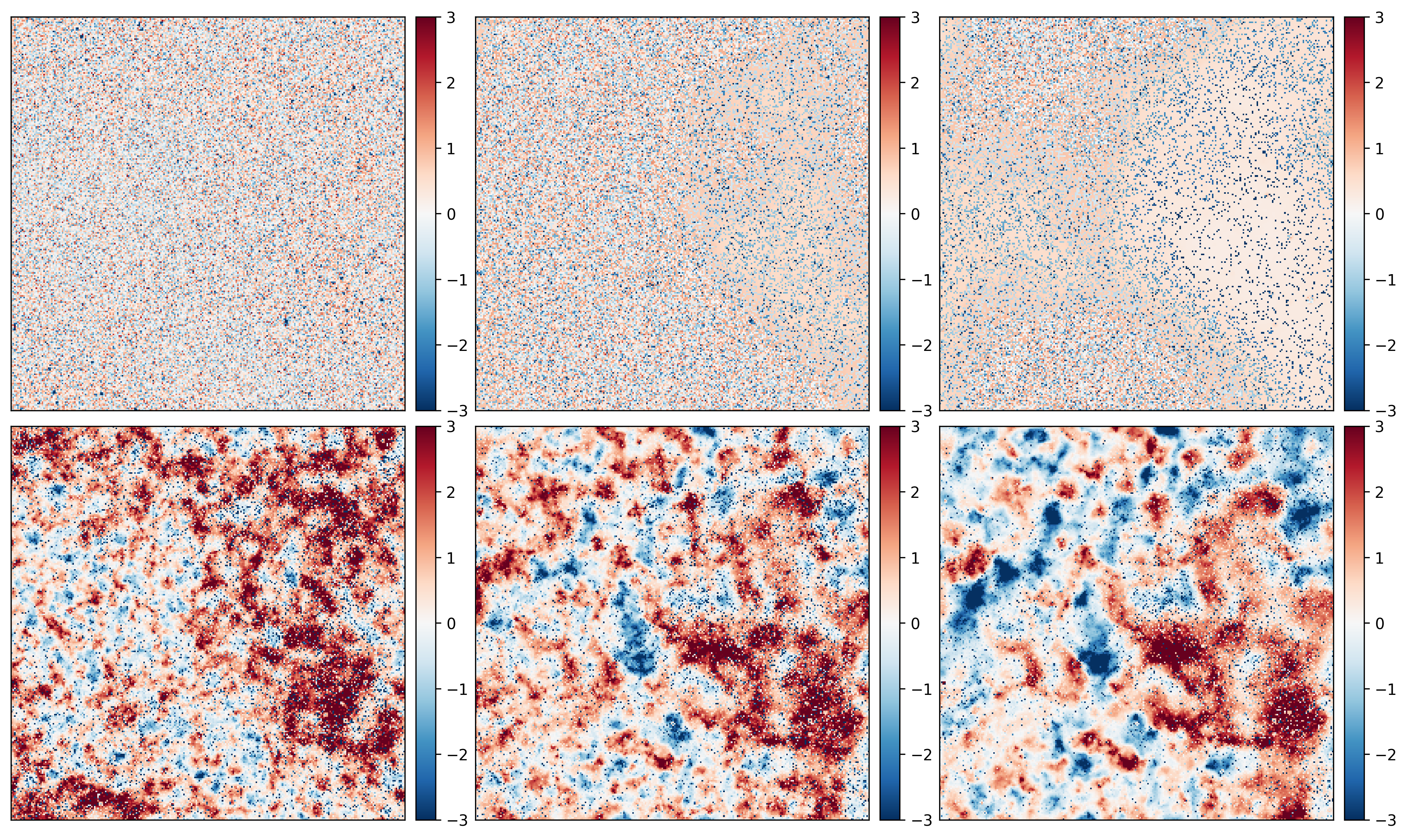}
\caption{
   Spectral reconstruction from the synthetic multi-frequency imaging example in Sec.~\ref{sec:mf_example}.
   \textit{Top row:} ground truth spectral index map $\alpha(\vb{x})$ (left) and spectral deviations $\log I_\delta(\vb{x}, \nu)$ as defined in Eq.~\eqref{eq:diffuse_mf}. 
   The central and right panels show spectral fluctuations at $\nu_\text{low}$ and $\nu_\text{hi}$, respectively. 
   \textit{Bottom row:} corresponding posterior mean reconstructions. These illustrate the method’s ability to recover both the large-scale spectral slope and localized deviations across the frequency range.
}
   \label{fig:3d_demo_residuals}
\end{figure*}

Figure~\ref{fig:3d_demo_residuals} shows nwr and uwr across three frequency bands for the multi-frequency example, confirming accurate reconstruction across the entire spectral range.

\subsection{LMC SN1987A eROSITA reconstruction}\label{app:erosita_diagnostics}
\begin{figure*}[!htbp]
   \centering
   \includegraphics[width=1.\textwidth]{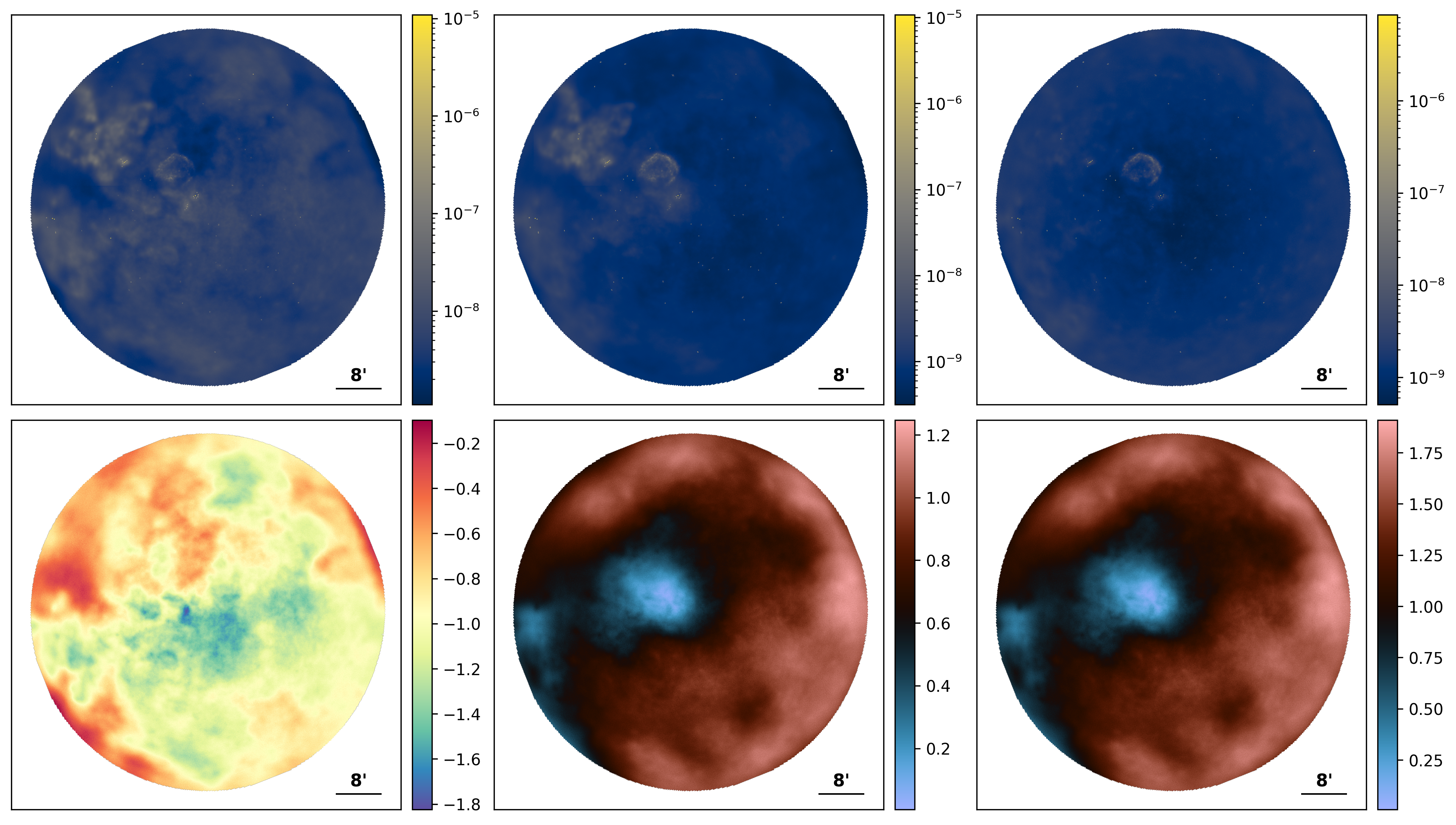}
    \caption{
	eROSITA LMC SN1987A reconstruction.
	The top row shows the reconstructed sky photon flux in the three energy bands, from left to right:
	$\nu_\text{low} = 0.2\text{ - }1.0,\mathrm{keV}$,
	$\nu_\text{ref} = 1.0\text{ - }2.0,\mathrm{keV}$, and
	$\nu_\text{hi} = 2.0\text{ - }4.5,\mathrm{keV}$.
	The bottom row displays the posterior mean of the reconstructed spectral index map $\alpha(\vb{x})$ and spectral deviation map $\log I_\delta(\vb{x}, \nu)$ (see Eq.~\eqref{eq:diffuse_mf}) for the diffuse background component.
    }
   \label{fig:erosita_spectral_3d}
\end{figure*}

\begin{figure*}[!htbp]
   \centering
   \includegraphics[width=1.\textwidth]{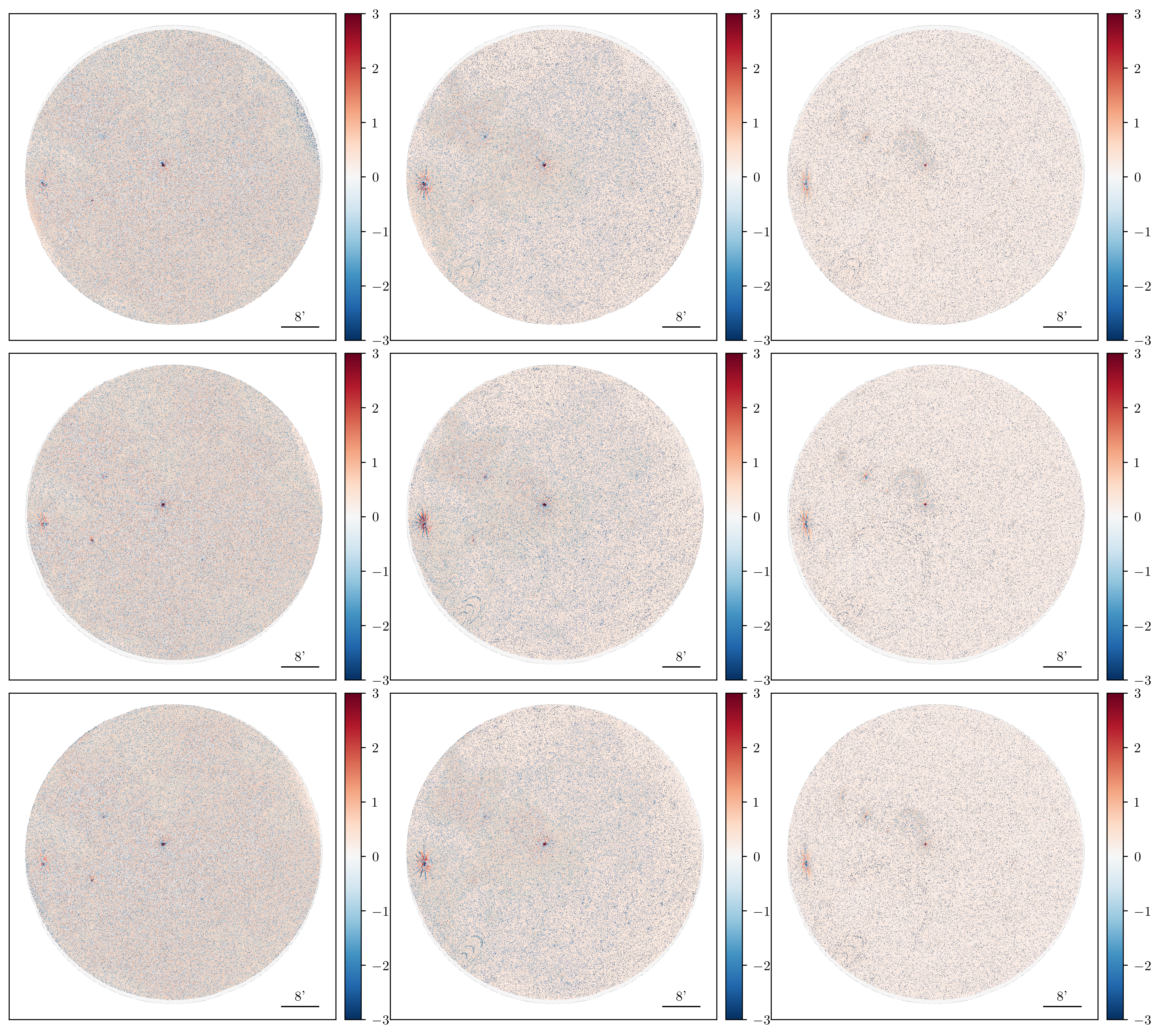}
    \caption{
	eROSITA LMC SN1987A data noise-weighted residuals.
	The columns represent, from left to right, the nwr in the three energy channels ($\nu_\text{low} = 0.2\text{ - }1.0,\mathrm{keV}$, $\nu_\text{ref} = 1.0\text{ - }2.0,\mathrm{keV}$, and $\nu_\text{hi} = 2.0\text{ - }4.5,\mathrm{keV}$), while the rows correspond 		to the three telescope modules used in the reconstruction (TM1, TM3, TM4).
    }
   \label{fig:erosita_nwr}
\end{figure*}

We present additional diagnostics of the eROSITA LMC SN1987A reconstruction.
Figure~\ref{fig:erosita_spectral_3d} shows the posterior mean of the reconstructed sky photon flux for each energy band, along with the posterior mean maps of the background spectral index and spectral deviations.
\begin{figure*}[tbp!]
   \centering
   \includegraphics[width=\textwidth]{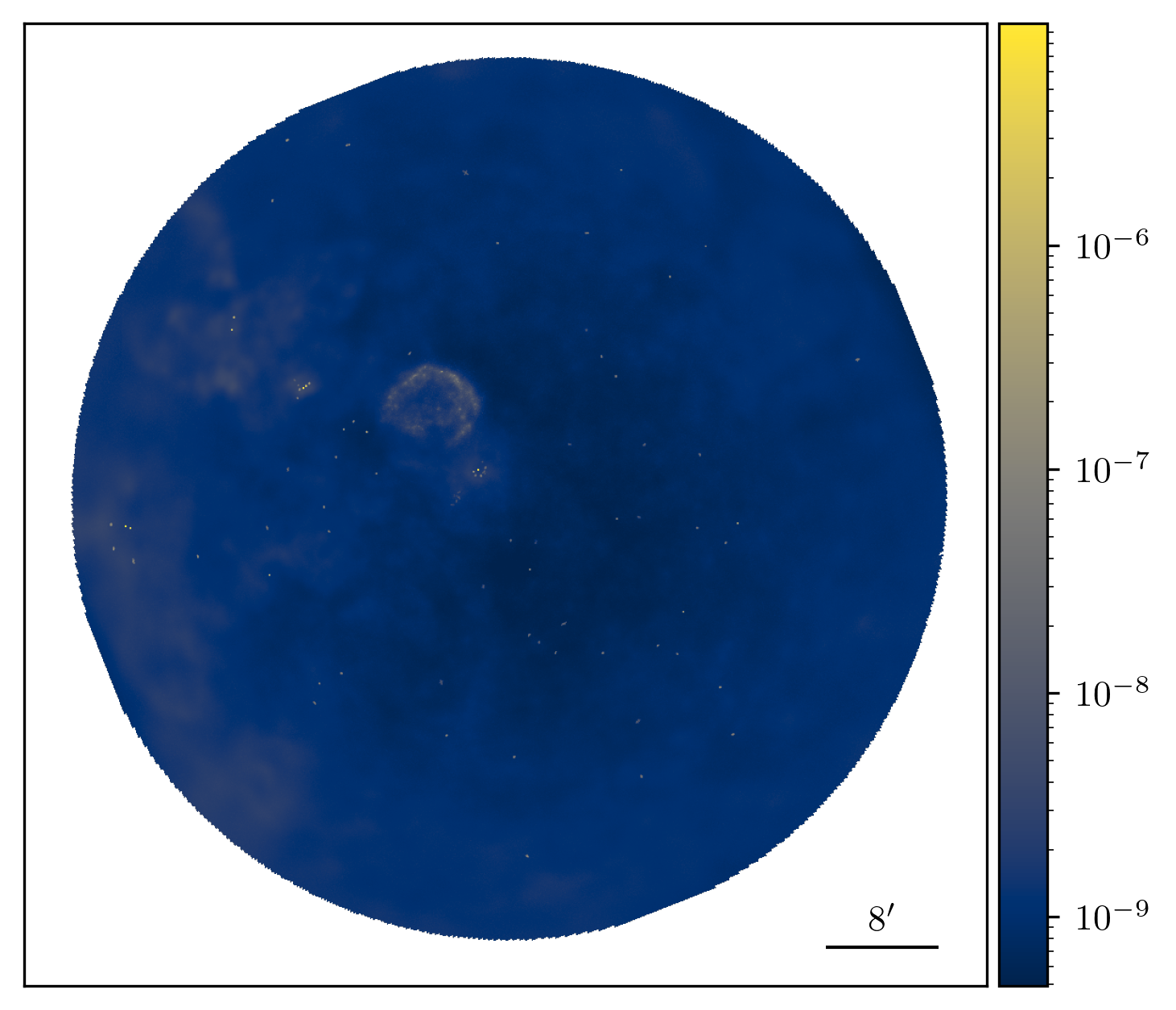}
    \caption{
   Posterior mean reconstruction of the eROSITA LMC SN1987A sky photon flux at an unobserved intermediate energy band (here, $2.4\,\mathrm{keV}$), 
   predicted from the joint multi-frequency model in Eq.~\eqref{eq:mf_sky}.
   This demonstrates the model’s ability to interpolate the sky signal across energy bands based on the inferred spectral correlation structure.
    }
   \label{fig:erosita_intermediate_wavelength}
\end{figure*}
Figure~\ref{fig:erosita_nwr} presents the corresponding noise-weighted residuals, computed according to Eq.~\eqref{eq:nwr}.
We also show the ability of the model to predict the photon flux distribution at unobserved wavelengths in Fig.~\ref{fig:erosita_intermediate_wavelength}.


Table~\ref{tab:point_sources} lists all detected point-like sources.
In a post-processing step, we removed all point-source candidates (48 in total) for which the posterior standard deviation of the flux density exceeded 1.5 times the posterior mean, or for which the positional uncertainty exceeded two pixels.
This procedure eliminates poorly constrained point-source components that do not meaningfully contribute to the reconstruction.
We emphasize that this list does not constitute a validated astrophysical source catalog.
The detection algorithm may flag sharp edges or residual structures -- often arising from limitations in the PSF model -- as spurious point sources.
These residuals primarily stem from incomplete or imperfect knowledge of the instrument, such as inaccuracies in the calibration database and corresponding response implementation.
For further discussion of these systematics, see \citet{eberle24arxiv}.
We visually inspected the detected sources and flagged those likely attributable to such systematics in the rightmost column of Table~\ref{tab:point_sources}.

\begin{table*}[!htbp]
\centering
\caption{Detected point sources in the SN1987A field. Positions are $x \pm \sigma_x$, $y \pm \sigma_y$ in pixel coordinates. Logarithmic photon flux densities are measured in \si{\per\second\per\arcsecond} and given with their associated 1$\sigma$ posterior uncertainties. A value of 1 in the final column indicates sources that are flagged manually as possible artifacts.}
\label{tab:point_sources}
\begin{tabular}{crrrc}
\toprule
\shortstack{Positions\\(px)} & \shortstack{log flux density\\\small($0.2\text{–}1$\,keV)} & \shortstack{log flux density\\\small($1\text{–}2$\,keV)} & \shortstack{log flux density\\\small($2\text{–}4.5$\,keV)} & \shortstack{Possible\\artifact} \\
\midrule
108.37 ± 0.01, \quad 488.84 ± 0.01 & -8.92 ± 0.01 & -8.40 ± 0.01 & -8.63 ± 0.01 & 0 \\
483.27 ± 0.01, \quad 549.33 ± 0.01 & -8.50 ± 0.00 & -9.17 ± 0.00 & -10.12 ± 0.01 & 0 \\
297.25 ± 0.05, \quad 635.97 ± 0.11 & -11.04 ± 0.08 & -10.34 ± 0.08 & -10.47 ± 0.08 & 0 \\
113.28 ± 0.09, \quad 486.88 ± 0.08 & -10.21 ± 0.03 & -11.78 ± 0.13 & -10.76 ± 0.09 & 0 \\
230.08 ± 0.15, \quad 892.54 ± 0.24 & -9.52 ± 0.05 & -11.12 ± 0.08 & -12.40 ± 0.24 & 0 \\
299.79 ± 0.14, \quad 638.38 ± 0.14 & -12.32 ± 0.18 & -11.45 ± 0.06 & -11.66 ± 0.24 & 0 \\
260.80 ± 0.02, \quad 437.37 ± 0.03 & -10.08 ± 0.02 & -12.01 ± 0.02 & -13.90 ± 0.21 & 0 \\
221.13 ± 0.05, \quad 698.05 ± 0.06 & -12.70 ± 0.14 & -12.02 ± 0.05 & -12.13 ± 0.09 & 0 \\
223.38 ± 0.10, \quad 711.39 ± 0.09 & -12.53 ± 0.10 & -12.17 ± 0.05 & -12.99 ± 0.21 & 0 \\
759.08 ± 0.05, \quad 492.39 ± 0.08 & -12.66 ± 0.04 & -13.45 ± 0.07 & -13.51 ± 0.11 & 0 \\
478.26 ± 0.14, \quad 546.88 ± 0.20 & -12.46 ± 0.13 & -13.25 ± 0.08 & -15.12 ± 0.39 & 0 \\
303.33 ± 0.31, \quad 640.88 ± 0.38 & -13.36 ± 0.18 & -13.42 ± 0.17 & -14.55 ± 0.76 & 0 \\
116.31 ± 0.30, \quad 451.96 ± 1.19 & -13.38 ± 0.35 & -13.66 ± 0.07 & -14.82 ± 0.36 & 1 \\
701.40 ± 0.04, \quad 397.79 ± 0.04 & -13.50 ± 0.09 & -14.59 ± 0.24 & -13.93 ± 0.21 & 0 \\
92.46 ± 0.33, \quad 490.79 ± 0.59 & -13.87 ± 0.50 & -13.27 ± 0.12 & -14.87 ± 0.70 & 1 \\
293.05 ± 0.34, \quad 634.29 ± 0.42 & -12.97 ± 0.25 & -13.44 ± 0.26 & -15.58 ± 0.36 & 0 \\
364.48 ± 0.08, \quad 589.42 ± 0.08 & -15.61 ± 0.37 & -13.53 ± 0.04 & -13.04 ± 0.06 & 0 \\
740.36 ± 0.27, \quad 317.60 ± 0.12 & -13.78 ± 0.21 & -14.35 ± 0.27 & -14.18 ± 0.29 & 0 \\
635.14 ± 0.33, \quad 867.83 ± 0.16 & -12.28 ± 0.06 & -14.36 ± 0.23 & -15.99 ± 0.95 & 0 \\
279.93 ± 0.68, \quad 899.75 ± 0.58 & -14.91 ± 0.35 & -14.28 ± 0.18 & -14.04 ± 0.42 & 0 \\
95.24 ± 0.24, \quad 465.33 ± 0.90 & -15.48 ± 1.61 & -13.28 ± 0.09 & -14.80 ± 0.89 & 1 \\
687.02 ± 0.21, \quad 754.48 ± 0.16 & -14.11 ± 0.19 & -14.49 ± 0.14 & -15.09 ± 0.54 & 0 \\
535.17 ± 0.62, \quad 137.99 ± 0.51 & -14.69 ± 0.74 & -14.26 ± 0.08 & -14.81 ± 0.62 & 0 \\
344.52 ± 1.29, \quad 894.66 ± 0.49 & -14.64 ± 0.53 & -14.53 ± 0.22 & -14.83 ± 0.50 & 0 \\
449.49 ± 0.25, \quad 266.39 ± 0.20 & -14.54 ± 0.14 & -14.88 ± 0.07 & -14.89 ± 0.22 & 0 \\
485.93 ± 0.27, \quad 542.76 ± 0.59 & -13.79 ± 0.33 & -14.75 ± 0.26 & -15.85 ± 0.64 & 1 \\
754.00 ± 0.59, \quad 267.49 ± 0.32 & -14.68 ± 0.44 & -14.77 ± 0.17 & -15.23 ± 0.60 & 0 \\
314.23 ± 0.18, \quad 321.66 ± 0.27 & -13.95 ± 0.29 & -14.93 ± 0.13 & -15.78 ± 0.60 & 0 \\
724.84 ± 0.22, \quad 786.80 ± 0.14 & -12.93 ± 0.09 & -14.97 ± 0.16 & -16.82 ± 0.53 & 0 \\
184.75 ± 0.23, \quad 456.68 ± 0.60 & -16.04 ± 0.61 & -14.58 ± 0.15 & -14.35 ± 0.47 & 0 \\
264.24 ± 0.16, \quad 835.43 ± 0.62 & -15.50 ± 0.93 & -14.75 ± 0.23 & -14.95 ± 0.68 & 0 \\
886.52 ± 0.98, \quad 666.21 ± 0.66 & -15.40 ± 0.77 & -14.92 ± 0.46 & -14.98 ± 0.40 & 0 \\
290.83 ± 0.34, \quad 625.30 ± 0.34 & -14.16 ± 0.29 & -14.82 ± 0.43 & -16.26 ± 0.47 & 0 \\
340.25 ± 0.13, \quad 592.13 ± 0.26 & -16.22 ± 0.46 & -14.85 ± 0.12 & -14.42 ± 0.20 & 0 \\
318.82 ± 0.21, \quad 509.42 ± 0.41 & -15.25 ± 0.15 & -15.14 ± 0.14 & -15.15 ± 0.32 & 0 \\
490.64 ± 0.33, \quad 551.07 ± 0.50 & -13.37 ± 0.25 & -15.14 ± 0.56 & -16.89 ± 0.92 & 1 \\
538.03 ± 0.19, \quad 443.07 ± 0.21 & -15.37 ± 0.40 & -15.37 ± 0.08 & -15.26 ± 0.35 & 0 \\
503.63 ± 0.39, \quad 790.24 ± 0.19 & -15.77 ± 0.71 & -15.16 ± 0.14 & -15.08 ± 0.51 & 0 \\
337.48 ± 0.50, \quad 332.43 ± 0.34 & -15.44 ± 0.72 & -15.21 ± 0.25 & -15.41 ± 0.60 & 0 \\
521.24 ± 0.34, \quad 243.54 ± 0.60 & -15.38 ± 0.70 & -15.23 ± 0.13 & -15.46 ± 0.52 & 0 \\
674.09 ± 0.36, \quad 361.87 ± 0.34 & -15.07 ± 0.43 & -15.46 ± 0.23 & -15.75 ± 0.52 & 0 \\
656.76 ± 0.46, \quad 222.89 ± 0.62 & -15.75 ± 0.90 & -15.27 ± 0.27 & -15.57 ± 0.58 & 0 \\
630.54 ± 0.25, \quad 496.85 ± 0.26 & -15.31 ± 0.42 & -15.57 ± 0.19 & -15.83 ± 0.54 & 0 \\
598.66 ± 0.99, \quad 800.86 ± 0.23 & -15.70 ± 0.51 & -15.21 ± 0.28 & -15.86 ± 0.64 & 0 \\
694.77 ± 0.29, \quad 353.21 ± 0.26 & -15.04 ± 0.50 & -15.72 ± 0.26 & -16.20 ± 0.74 & 0 \\
291.61 ± 0.38, \quad 639.01 ± 0.20 & -13.37 ± 0.18 & -15.95 ± 0.05 & -17.50 ± 0.52 & 0 \\
308.93 ± 0.62, \quad 300.97 ± 0.58 & -16.08 ± 1.17 & -15.33 ± 0.39 & -15.74 ± 0.50 & 0 \\
615.64 ± 0.30, \quad 354.15 ± 0.32 & -16.20 ± 0.91 & -15.72 ± 0.45 & -15.32 ± 0.44 & 0 \\
258.47 ± 0.21, \quad 487.28 ± 0.94 & -15.92 ± 0.59 & -15.57 ± 0.17 & -15.79 ± 0.61 & 0 \\
479.93 ± 0.47, \quad 542.94 ± 0.63 & -13.83 ± 0.41 & -15.57 ± 1.01 & -17.67 ± 2.04 & 1 \\
489.71 ± 1.05, \quad 545.97 ± 1.06 & -13.94 ± 0.43 & -15.87 ± 0.87 & -17.44 ± 1.26 & 1 \\
374.77 ± 0.34, \quad 545.10 ± 0.21 & -15.89 ± 0.40 & -15.74 ± 0.36 & -15.82 ± 0.19 & 0 \\
614.23 ± 0.26, \quad 669.74 ± 0.58 & -15.88 ± 0.58 & -15.81 ± 0.23 & -15.76 ± 0.52 & 0 \\
350.40 ± 0.23, \quad 600.40 ± 0.51 & -16.69 ± 0.69 & -15.55 ± 0.15 & -15.22 ± 0.53 & 0 \\
746.26 ± 0.29, \quad 471.30 ± 0.52 & -16.35 ± 0.83 & -15.72 ± 0.12 & -15.61 ± 0.50 & 0 \\
469.59 ± 1.40, \quad 865.06 ± 1.28 & -16.22 ± 1.13 & -15.77 ± 0.41 & -15.96 ± 0.86 & 0 \\
324.28 ± 0.36, \quad 483.49 ± 0.29 & -17.04 ± 1.42 & -15.81 ± 0.22 & -15.62 ± 0.45 & 0 \\
\bottomrule
\end{tabular}
\end{table*}

\begin{table*}[!htbp]
\centering
\caption{Continuation of Tab.~\ref{tab:point_sources}.}
\label{tab:point_sources_2}
\begin{tabular}{crrrc}
\toprule
\shortstack{Positions\\(px)} & \shortstack{log flux density\\\small($0.2\text{–}1$\,keV)} & \shortstack{log flux density\\\small($1\text{–}2$\,keV)} & \shortstack{log flux density\\\small($2\text{–}4.5$\,keV)} & \shortstack{Possible\\artifact} \\
\midrule
629.44 ± 0.52, \quad 648.48 ± 0.64 & -17.10 ± 1.06 & -15.88 ± 0.29 & -15.66 ± 0.72 & 0 \\
287.73 ± 0.39, \quad 644.59 ± 0.39 & -13.44 ± 0.36 & -16.56 ± 0.92 & -18.41 ± 1.27 & 0 \\
718.66 ± 0.33, \quad 565.17 ± 0.69 & -16.74 ± 1.27 & -15.90 ± 0.09 & -16.27 ± 0.78 & 0 \\
715.98 ± 0.42, \quad 487.18 ± 0.29 & -17.20 ± 1.46 & -15.84 ± 0.09 & -15.89 ± 0.51 & 0 \\
517.54 ± 0.27, \quad 473.98 ± 0.45 & -16.34 ± 0.84 & -16.19 ± 0.16 & -16.48 ± 0.56 & 0 \\
419.68 ± 0.38, \quad 595.41 ± 0.18 & -13.58 ± 0.20 & -16.69 ± 0.34 & -18.70 ± 0.90 & 0 \\
654.17 ± 0.90, \quad 498.55 ± 0.48 & -15.15 ± 0.54 & -16.69 ± 0.35 & -17.52 ± 0.54 & 0 \\
653.14 ± 1.17, \quad 281.40 ± 1.08 & -16.68 ± 0.96 & -16.29 ± 0.35 & -16.61 ± 0.89 & 0 \\
331.73 ± 0.28, \quad 562.60 ± 0.45 & -17.77 ± 1.52 & -16.28 ± 0.49 & -15.54 ± 0.57 & 0 \\
580.55 ± 0.94, \quad 576.01 ± 0.44 & -15.13 ± 0.33 & -16.70 ± 0.43 & -17.89 ± 0.87 & 0 \\
410.25 ± 0.48, \quad 672.95 ± 0.64 & -18.12 ± 1.91 & -16.17 ± 0.40 & -15.49 ± 0.48 & 0 \\
462.50 ± 0.61, \quad 519.71 ± 0.80 & -14.24 ± 0.73 & -16.70 ± 0.38 & -18.69 ± 1.32 & 0 \\
537.24 ± 0.50, \quad 373.28 ± 0.55 & -17.56 ± 1.56 & -16.25 ± 0.30 & -16.15 ± 0.43 & 0 \\
565.37 ± 0.21, \quad 354.42 ± 0.77 & -17.55 ± 1.52 & -16.00 ± 0.13 & -16.42 ± 0.91 & 0 \\
459.84 ± 0.18, \quad 515.75 ± 0.67 & -13.95 ± 0.19 & -16.91 ± 0.57 & -18.89 ± 1.02 & 0 \\
443.61 ± 0.73, \quad 323.36 ± 1.24 & -16.59 ± 0.74 & -16.69 ± 0.64 & -16.92 ± 0.96 & 0 \\
280.62 ± 0.41, \quad 549.14 ± 0.91 & -17.74 ± 1.69 & -16.71 ± 0.74 & -15.75 ± 0.50 & 0 \\
462.26 ± 0.82, \quad 525.21 ± 0.47 & -14.05 ± 0.34 & -17.01 ± 0.62 & -19.13 ± 1.29 & 0 \\
660.01 ± 0.77, \quad 575.79 ± 0.59 & -17.06 ± 0.91 & -16.69 ± 0.63 & -16.87 ± 0.59 & 0 \\
574.04 ± 1.40, \quad 385.20 ± 0.80 & -18.04 ± 1.04 & -16.99 ± 0.55 & -15.53 ± 0.42 & 0 \\
487.49 ± 0.26, \quad 557.36 ± 0.26 & -14.03 ± 0.29 & -17.25 ± 0.59 & -19.06 ± 1.00 & 1 \\
488.25 ± 0.68, \quad 424.83 ± 0.92 & -16.72 ± 0.66 & -17.02 ± 0.35 & -17.16 ± 0.87 & 0 \\
547.73 ± 0.56, \quad 365.26 ± 0.65 & -17.00 ± 1.31 & -16.86 ± 0.61 & -17.18 ± 0.78 & 0 \\
455.83 ± 1.06, \quad 511.60 ± 0.51 & -15.17 ± 0.63 & -17.22 ± 0.38 & -18.84 ± 0.83 & 0 \\
459.63 ± 1.76, \quad 521.05 ± 0.92 & -14.81 ± 0.97 & -17.34 ± 0.57 & -19.30 ± 1.70 & 0 \\
544.68 ± 0.62, \quad 472.26 ± 1.10 & -14.98 ± 0.35 & -17.59 ± 0.95 & -19.08 ± 1.31 & 0 \\
597.97 ± 0.58, \quad 697.99 ± 0.61 & -17.03 ± 1.22 & -17.51 ± 0.97 & -17.98 ± 0.80 & 0 \\
\bottomrule
\end{tabular}
\end{table*}

\section{Slope deviations degeneracy}\label{app:slope_removal}

The logarithm of the spectral deviations term $\log{I_\delta(\vb{x}, \nu)}$ from Eq.~\eqref{eq:frequency_deviations} reads
\begin{equation*}
	\log I_\delta(\vb{x}, \nu) = \tilde{\delta}(\vb{x}, \nu) - \bar{\delta}(\vb{x}, \nu) \log{\frac{\nu}{\nu_\text{ref}}}.
\end{equation*}
Suppose that the deviation field $\tilde{\delta}(\vb{x}, \nu)$ contains a slope-like term in addition to residual structure
\begin{equation*}
    \tilde{\delta}(\vb{x}, \nu) \coloneqq \alpha'(\vb{x})\, \log{\frac{\nu}{\nu_\text{ref}}} + \tilde{\delta}_R(\vb{x}, \nu),
\end{equation*}
where $\alpha'(\vb{x})$ is a spurious spectral slope introduced by the stochastic deviation process, and $\tilde{\delta}_R(\vb{x}, \nu)$ captures variations orthogonal to the power-law subspace.
The term $\alpha'(\vb{x}) \log(\nu/\nu_\text{ref})$ is degenerate with the main power-law component parametrized by $\alpha(\vb{x}) \log(\nu/\nu_\text{ref})$ in Eq.~\eqref{eq:frequency_deviations}. 

To eliminate this degeneracy, we subtract the empirical average slope $\bar{\delta}(\vb{x})$ defined in Eq.~\eqref{eq:slope_removal}, which removes the power-law-like component in $\tilde{\delta}(\vb{x}, \nu)$
\begin{equation*}
\begin{aligned}
    \bar{\delta}(\vb{x}) &= 
    \frac{
        \int_{\Omega_\nu}
        \delta(\vb{x}, \tilde{\nu})\, \log{\frac{\tilde{\nu}}{\nu_\text{ref}}}\, \dd{\tilde{\nu}}
    }{
        \int_{\Omega_\nu}
        \qty(\log{\frac{\tilde{\nu}}{\nu_\text{ref}}})^2\, \dd{\tilde{\nu}}
    } \\
    &=
    \alpha'(\vb{x}) 
    + \frac{
        \int_{\Omega_\nu}
        \tilde{\delta}_R(\vb{x}, \tilde{\nu})\, \log{\frac{\tilde{\nu}}{\nu_\text{ref}}}\, \dd{\tilde{\nu}}
    }{
        \int_{\Omega_\nu}
        \qty(\log{\frac{\tilde{\nu}}{\nu_\text{ref}}})^2\, \dd{\tilde{\nu}}
    } \\
    &\eqqcolon \alpha'(\vb{x}) + \alpha_{R}(\vb{x}).
\end{aligned}
\end{equation*}
The resulting expression for $\tilde{\delta}(\vb{x}, \nu)$ thus becomes
\begin{equation*}
	\tilde{\delta}(\vb{x}, \nu) = \tilde{\delta}_R(\vb{x}, \nu) - \alpha_{R}(\vb{x})\, \log{\frac{\nu}{\nu_\text{ref}}}.
\end{equation*}
This formulation ensures that any spectral slope contained in the stochastic deviation field is projected out, preserving the interpretability and identifiability of the true spectral index $\alpha(\vb{x})$ in Eq.~\eqref{eq:frequency_deviations}. 
In the ideal case where $\tilde{\delta}_R(\vb{x}, \nu)$ contains no power-law-like contribution, the correction term $\alpha_R(\vb{x})$ vanishes, and $\bar{\delta}(\vb{x}) = \alpha'(\vb{x})$ as expected.
\end{appendix}
\end{document}